\documentclass[preprint,nofootinbib]{revtex4}

\usepackage{amsmath}
\usepackage{graphicx}
\usepackage[bookmarks=false]{hyperref} 

\newcommand{\beq}{\begin{equation}}
\newcommand{\eeq}{\end{equation}}
\newcommand{\beqa}{\begin{eqnarray}}
\newcommand{\eeqa}{\end{eqnarray}}
\newcommand{\no}{\nonumber}
\def\OMIT#1{{}}
\newcommand{\lsim}{\mathrel{\rlap{\lower4pt\hbox{\hskip1pt$\sim$}}
    \raise1pt\hbox{$<$}}}         
\newcommand{\gsim}{\mathrel{\rlap{\lower4pt\hbox{\hskip1pt$\sim$}}
    \raise1pt\hbox{$>$}}}         

\begin{document}
\begin{flushright}
UB-ECM-PF-09/17\\
ICCUB-09-213\\
PI/UAN2009-344FT\\
\end{flushright}

\vspace*{.5cm}

\title{BMSSM Implications for Cosmology}

\author{Nicol\'as Bernal}
\affiliation{High Energy Physics Group, Dept ECM and Institut de
  Ci\`encies del Cosmos,\\  Universitat de Barcelona, Av. Diagonal
  647, E-08028, Barcelona, Catalonia, Spain}
\author{Kfir Blum}
\affiliation{Department of Particle Physics, Weizmann
  Institute of Science, Rehovot 76100, Israel}
\author{Marta Losada}
\affiliation{Centro de Investigaciones, Cra 3 Este No 47A-15,
  Universidad Antonio Nari\~{n}o, Bogot\'a, Colombia}
\author{Yosef Nir}
\affiliation{Department of Particle Physics, Weizmann Institute of
  Science, Rehovot 76100, Israel}

\begin{abstract}
  The addition of non-renormalizable terms involving the Higgs fields
  to the MSSM (BMSSM) ameliorates the little hierarchy problem of the
  MSSM. We analyze in detail the two main cosmological issues affected
  by the BMSSM: dark matter and baryogenesis. The regions for which
  the relic abundance of the LSP is consistent with WMAP and collider
  constraints are identified, showing that the bulk region and other
  previously excluded regions are now permitted. Requiring vacuum
  stability limits the allowed regions. Based on a two-loop finite
  temperature effective potential analysis, we show that the
  electroweak phase transition can be sufficiently first order in
  regions that for the MSSM are incompatible with the LEP Higgs mass
  bound, including parameter values of $\tan\beta \lsim 5$,
  $m_{\tilde{t}_{1}} > m_t$, $m_Q \ll $ TeV.
\end{abstract}

\maketitle
\section{Introduction}
\label{sec:intro}
The smallness of the quartic Higgs coupling in the framework of the
minimal supersymmetric standard model (MSSM) poses a problem. The tree
level bound on the Higgs mass is violated, and
large enough loop corrections to satisfy the lower bound on the Higgs
mass suggest that the stop sector has rather peculiar features:  at
least one of the stop mass eigenstates should be rather heavy and/or
left-right-stop mixing should be substantial.

The situation is different if the quartic Higgs couplings are affected
by new physics. If the new physics appears at an energy scale that is
somewhat higher than the electroweak breaking scale, then its effects
can be parameterized by non-renormalizable terms. The leading
non-renormalizable terms that modify the quartic couplings are
\cite{Strumia:1999jm,Brignole:2003cm,Casas:2003jx,PRS,Dine:2007xi,Batra:2008rc,Antoniadis:2008es}:
\beq\label{eq:wdst}
W_{\rm BMSSM}=\frac{\lambda_1}{M}(H_uH_d)^2+\frac{\lambda_2}{M}{\cal
  Z}(H_uH_d)^2,
\eeq
where ${\cal Z}$ is a SUSY-breaking spurion:
\beq
{\cal Z}=\theta^2m_{\rm susy}.
\eeq
The first term in Eq. (\ref{eq:wdst}) is supersymmetric, while the
second breaks supersymmetry (SUSY). In the scalar potential, the
following quartic terms are generated:
\beq\label{eq:dstsp}
2\epsilon_1 H_uH_d(H_u^\dagger H_u+H_d^\dagger H_d)
+\epsilon_2(H_uH_d)^2,
\eeq
where
\beq
\epsilon_1\equiv\frac{\mu^*\lambda_1}{M},\ \ \
\ \ \ \epsilon_2\equiv-\frac{m_{\rm susy}\lambda_2}{M}.
\eeq

The interplay between the Higgs sector, the stop sector, and the
non-renormalizable (NR) operators has interesting consequences for the
MSSM baryogenesis \cite{Blum:2008ym}. The window for MSSM baryogenesis
is extended and, most important, can be made significantly more
natural. In addition, these operators have implications for yet another
cosmological issue, and that is dark matter \cite{Cheung:2009qk}.
In this work we present an extended analysis of the results for both the
electroweak phase transition and the dark matter relic abundance in the BMSSM.

One of the attractive features of the MSSM is the fact that the
lightest R-parity-odd particle (LSP)
is a natural candidate for being the dark matter particle.
Progress in experimentally constraining the MSSM parameter space
restricts, however, the regions where the dark matter is
quantitatively accounted for to rather special regions of the MSSM:
the focus point region, with surprisingly heavy sfermions; the funnel
region, where the mass of the CP-odd neutral Higgs scalar is very
close to twice the mass of the LSP; the co-annihilation region, where
the mass of the scalar partner of the right-handed tau is very close
to the mass of the LSP; and the bulk region, where the bino-LSP and
the sleptons are light.

The effects of the NR operators are potentially important for two of
these four regions. First, these operators give rise to a new
Higgs-Higgs-higgsino-higgsino interaction Lagrangian,
\beq\label{eq:hhHH}
-\frac{\epsilon_1}{\mu^*}\left[2(H_uH_d)(\widetilde H_u\widetilde H_d)
+2(\widetilde H_u H_d)(H_u\widetilde H_d)+(H_u\widetilde H_d)
(H_u\widetilde H_d)+(\widetilde H_u H_d)(\widetilde H_u H_d)\right]
+{\rm h.c.},
\eeq
which contributes to the annihilation process of two higgsinos to two
Higgs particles. This effect is relevant when the dark matter particle
has a significant component of higgsinos, as is the case in the focus
point region. Second, as mentioned above, these operators modify the
relation between the light Higgs mass and the stop masses. This effect
can be important in the bulk region within models where the slepton and
stop masses are related, such as the mSUGRA models. In this work, we will
study these effects and assess their quantitative significance.

The plan of this paper is as follows. In Section \ref{sec:spectrum} we
present the BMSSM spectra of the Higgs, neutralino and chargino sectors,
and the implications for the stop spectrum. In Section \ref{sec:ann} we
describe the BMSSM modifications to the annihilation cross sections that
are relevant to the dark matter relic abundance. In Section \ref{sec:dmrel}
we analyze the implications of the BMSSM operators for dark matter, while
in Section \ref{sec:ewpt} we explore the parameter space where the
electroweak phase transition (EWPT) is strongly first order, as required
for successful baryogenesis. We summarize our conclusions in Section
\ref{sec:con}.

\section{The spectrum}
\label{sec:spectrum}

\subsection{The Higgs sector and light stops}
We define the scalar Higgs components by
\beqa
H_d&=&\begin{pmatrix} H_d^0\\ H_d^- \end{pmatrix}
=\begin{pmatrix}\frac{\phi_1+H_{dr}+iH_{di}}{\sqrt2}\\ H_d^-
\end{pmatrix}\no\\
H_u&=&\begin{pmatrix} H_u^+\\ H_u^0 \end{pmatrix}
=\begin{pmatrix}H_u^+\\ \frac{\phi_2+H_{ur}+iH_{ui}}{\sqrt2}
\end{pmatrix}.
\eeqa
The VEVs of these Higgs fields are parameterized by
\beqa
\langle H_d^0\rangle&=&\phi_1/\sqrt{2},\ \ \ \langle
H_u^0\rangle=\phi_2/\sqrt{2},\\
\tan\beta&=&|\phi_2/\phi_1|,\ \ \ v=\sqrt{(\phi_1^2+\phi_2^2)/2}\simeq
174\text{ GeV}.\no
\eeqa
To leading order, the two charged and four neutral Higgs mass
eigenstates are related to the interaction eigenstates via
\beqa\label{eq:Hdef}
\begin{pmatrix} H_d^{*+}\\ H_u^+ \end{pmatrix}&=&
\begin{pmatrix} s_\beta & -c_\beta \\ c_\beta & s_\beta \end{pmatrix}
\begin{pmatrix} H^+ \\ G^+ \end{pmatrix},\no\\
\begin{pmatrix} H_{di}\\ H_{ui} \end{pmatrix}&=&
\begin{pmatrix} s_\beta & -c_\beta \\ c_\beta & s_\beta \end{pmatrix}
\begin{pmatrix} A \\ G^0 \end{pmatrix},\no\\
\begin{pmatrix} H_{dr}\\ H_{ur} \end{pmatrix}&=&
\begin{pmatrix} c_\alpha & -s_\alpha \\ s_\alpha & c_\alpha
\end{pmatrix}
\begin{pmatrix} H \\ h \end{pmatrix},
\eeqa
where $c_\beta\equiv\cos\beta$, $s_\beta\equiv\sin\beta$, and
similarly for $\alpha$. Within the MSSM (without the $\epsilon_i$
operators), the angle $\alpha$ is given by (at tree level)
\beq\label{alphamssm} s_{2\alpha}=-\frac{m_A^2+m_Z^2}{m_H^2-m_h^2}\
s_{2\beta}. \eeq If the $\epsilon_{1,2}$ couplings are complex, then
the four neutral mass eigenstates are related by a $4\times 4$
transformation matrix to the real and imaginary components of $H_d^0$
and $H_u^0$. The analysis of dark matter is, however, unaffected at
leading order, so we neglect such effects here. In the unitary gauge,
the Goldstone fields $G^\pm$ and $G^0$ are set to zero.

Taking $m_Z,m_A$ and $\tan\beta$ as input parameters, we obtain the
following $\epsilon_i$-corrections to the Higgs spectrum:
\beqa\label{eq:delm}
\delta_\epsilon m_h^2&=&2v^2\left(\epsilon_{2r}-2\epsilon_{1r}s_{2\beta}
-\frac{2\epsilon_{1r}(m_A^2+m_Z^2)s_{2\beta}
+\epsilon_{2r}(m_A^2-m_Z^2)c^2_{2\beta}}
{\sqrt{(m_A^2-m_Z^2)^2+4m_A^2m_Z^2s^2_{2\beta}}}\right),\no\\
\delta_\epsilon m_H^2&=&2v^2\left(\epsilon_{2r}-2\epsilon_{1r}s_{2\beta}
+\frac{2\epsilon_{1r}(m_A^2+m_Z^2)s_{2\beta}
+\epsilon_{2r}(m_A^2-m_Z^2)c^2_{2\beta}}
{\sqrt{(m_A^2-m_Z^2)^2+4m_A^2m_Z^2s^2_{2\beta}}}\right),\no\\
\delta_\epsilon m_{H^\pm}^2&=&2v^2\epsilon_{2r}.
\eeqa
The angle $\alpha$ is shifted from its MSSM value:
\beqa\label{eq:alpha}
s_{2\alpha}&=&\frac{-(m_A^2+m_Z^2)s_{2\beta}+4v^2\epsilon_{1r}}
{(m_H^2-m_h^2)s_{2\beta}}\\
&=&-\frac{(m_A^2+m_Z^2)s_{2\beta}}
{(m_A^4-2m_A^2m_Z^2c_{4\beta}+m_Z^4)^{1/2}}
-4v^2c_{2\beta}^2\frac{2\epsilon_{1r}(m_A^2-m_Z^2)^2
-\epsilon_{2r}s_{2\beta}(m_A^4-m_Z^4)}
{(m_A^4-2m_A^2m_Z^2c_{4\beta}+m_Z^4)^{3/2}}.\no
\eeqa

The possibility of a light Higgs scalar hidden in the LEP data is not
excluded (see e.g. \cite{lightH}) and, in fact, may have interesting
implications for DM~\cite{lightH_DM}. This scenario relies upon
sizable mixing in the Higgs mass matrix, such that production via
$e^+e^-\to Zh$ is suppressed. Hence the mass splitting in the scalar
Higgs sector must be rather small. A light Higgs sector is an
interesting possibility also within the BMSSM, following the impact of
the non-renormalizable operators on the mixing angles, captured to
leading order in Eq.~(\ref{eq:alpha}). An extreme example for this
possibility was presented in \cite{Batra:2008rc}, in which the
electroweak symmetry breaking vacuum is controlled by the
non-renormalizable operators.

In the present work, we limit our attention to the more conservative
situation wherein the non-renormalizable operators can still be
treated as perturbations in the usual electroweak breaking vacuum of
the MSSM. We allow for significant splitting between the heavy and
light Higgs mass eigenstates, in which case the constraint that arises
from the LEP bound on the light Higgs mass can be written as follows:
\beq\label{lowhig} m_h^2\approx m_h^{2({\rm
    tree})}+\delta _t m_h^2+\delta_\epsilon m_h^2 \gsim\big(114\text{
  GeV}\big)^2, \eeq
where
\beqa\label{eq:demh}
\delta _t m_h^2&=&\frac{3m_t^4}{4\pi^2 v^2}\ln\left(\frac
{m_{\tilde t_1} m_{\tilde t_2}}{m_t^2}\right)\no\\
&+&\frac{3m_t^4}{4\pi^2 v^2}\frac{|X_t|^2}{m_{\tilde t_1}^2
  -m_{\tilde t_2}^2}
\left[\ln\left(\frac{m_{\tilde t_1}^2}{m_{\tilde t_2}^2}\right)
+\frac12\frac{|X_t|^2}{m_{\tilde t_1}^2-m_{\tilde t_2}^2}\left(
2-\frac{m_{\tilde t_1}^2+m_{\tilde t_2}^2}{m_{\tilde t_1}^2-m_{\tilde t_2}^2}
\ln\left(\frac{m_{\tilde t_1}^2}{m_{\tilde t_2}^2}\right)\right)\right],\no\\
X_t&=&A_t+\mu^*\cot\beta.
\eeqa

The $\delta_\epsilon m_h^2$ contribution relaxes the constraints on
$\delta_t m_h^2$ in a significant way. In fact, $\delta_t m_h^2\leq0$
is not excluded (in this respect we disagree with the conclusions of
\cite{Antoniadis:2008es}).  Thus, for $\epsilon_1\lsim-0.05$ and
$\tan\beta\lsim10$, the two stop mass eigenstates can be as light as
the top quark, or one could be as light as the direct experimental
lower bound with the other only slightly heavier than the top.

\subsection{Neutralinos}
The neutralino mass matrix is given  by
\beq
M_{\widetilde N}=\begin{pmatrix}
M_1 & 0 & -m_Zs_Wc_\beta & m_Zs_Ws_\beta \\
0 & M_2 & m_Zc_Wc_\beta & -m_Zc_Ws_\beta \\
-m_Zs_Wc_\beta & m_Zc_Wc_\beta & 0 & -\mu \\
m_Zs_Ws_\beta & -m_Zc_Ws_\beta & -\mu & 0 \end{pmatrix}
+\frac{4\epsilon_1 m_W^2}{\mu^* g^2}
\begin{pmatrix} 0 & 0 & 0 & 0 \\ 0 & 0 & 0 & 0 \\
0 & 0 & s^2_\beta & s_{2\beta} \\ 0 & 0 & s_{2\beta} & c^2_\beta
\end{pmatrix}.
\eeq
The transition to the neutralino mass basis is obtained with a unitary
matrix $Z$:
\beq
M_{\widetilde N}=Z^T\ {\rm diag}(m_{\widetilde N_1},
m_{\widetilde N_2}, m_{\widetilde N_3},m_{\widetilde N_4})\ Z.
\eeq

The gaugino fraction in the LSP is defined as
\beq
R_{\tilde\lambda}=|Z_{11}|^2+|Z_{12}|^2,
\eeq
while the higgsino fraction is given by
$1-R_{\tilde\lambda}= |Z_{13}|^2+|Z_{14}|^2$.

\subsection{Charginos}
The chargino mass matrix is given  by
\beq
M_{\widetilde C}=\begin{pmatrix}
M_2 & \sqrt{2}m_Ws_\beta \\
\sqrt{2}m_Wc_\beta & \mu \end{pmatrix}
 -\frac{2\epsilon_1 m_W^2 s_{2\beta}}{\mu^* g^2}
\begin{pmatrix} 0 & 0 \\ 0 & 1 \end{pmatrix}.
\eeq
The transition to the neutralino mass basis is obtained with unitary
matrices $V$ and $U$:
\beq
M_{\widetilde C}=U^T\ {\rm diag}(m_{\widetilde C_1}, m_{\widetilde
  C_2})\ V.
\eeq

\section{Annihilation cross sections}
\label{sec:ann}
The effects of the BMSSM operators on (co)annihilation cross sections
are relevant to the dark matter issue when the LSP has a significant
component of higgsino, namely $1-R_{\tilde\lambda}\not\ll1$. These
effects come in two ways. First, there is an indirect effect, due to
the modification of the neutralino and chargino spectra. Since
co-annihilation rates are very sensitive to mass splittings, this is
often the more important effect. Second, there is a direct effect of
the new $\epsilon_1$-dependent couplings which modify the
(co)annihilation processes that involve Higgs scalars as mediators
and/or as final states. In this section, we focus on the latter
effect. In Section \ref{sec:dmrel} we analyze numerically the DM relic
abundance in the BMSSM, taking into account all effects.

\subsection{Single scalar}
\label{ssec:1h}
We consider terms of the form
\beq
C_\phi \phi\overline{\tilde N}\tilde N;\ \ \
C_\phi \phi\overline{\tilde N}\tilde C^+,\ \ \ \phi=h,H,A;H^-.
\eeq
We denote the MSSM couplings by $C_\phi^0$ and define the modification
that is induced by the $\epsilon_1$ terms as follows:
\beq
C_\phi=C_\phi^0(1-\delta_{\epsilon\phi}).
\eeq

As concerns the emission of a single neutral scalar in the
annihilation of two neutralino LSPs, we obtain the following
$\delta_{\epsilon\phi}$'s:
\beqa
\delta_{\epsilon h}&=&\frac{2\sqrt{2}\lambda_1^* v}{gM}\times
\frac{-c_\beta s_\alpha Z_{14}^2 +s_\beta c_\alpha
  Z_{13}^2+2c_{(\alpha+\beta)}Z_{13}Z_{14}}
{(Z_{12}-\tan\theta_W Z_{11})(s_\alpha Z_{13}+c_\alpha Z_{14})},\\
\delta_{\epsilon H}&=&\frac{2\sqrt{2}\lambda_1^* v}{gM}\times
\frac{c_\beta c_\alpha Z_{14}^2 +s_\beta s_\alpha
  Z_{13}^2+2s_{(\alpha+\beta)}Z_{13}Z_{14}}
{(Z_{12}-\tan\theta_W Z_{11})(-c_\alpha Z_{13}+s_\alpha Z_{14})},\\
\delta_{\epsilon A}&=&\frac{\sqrt{2}\lambda_1^* v}{gM}\times
\frac{s_{2\beta} (Z_{14}^2+Z_{13}^2)+4Z_{13}Z_{14}}
{(Z_{12}-\tan\theta_W Z_{11})(s_\beta Z_{13}-c_\beta Z_{14})}.
\eeqa

As concerns the emission of a single charged scalar in the
co-annihilation of the neutralino LSP with the lightest chargino,
we obtain the following $\delta_{\epsilon H^-}$'s:
\beqa
\delta_{\epsilon H^- R}&=&\frac{2\lambda_1^* v}{gM\tan\beta}\times
\frac{U_{12}(c_{\beta}Z_{14}+s_\beta Z_{13})}
{U_{11}Z_{13}-\frac{1}{\sqrt{2}}U_{12}(Z_{12}+\tan\theta_W Z_{11})},\\
\delta^*_{\epsilon H^- L}&=&\frac{2\lambda_1^* v\tan\beta}{gM}\times
\frac{V_{12}(c_{\beta}Z_{14}+s_\beta Z_{13})}
{V_{11}Z_{14}+\frac{1}{\sqrt{2}}V_{12}(Z_{12}+\tan\theta_W Z_{11})}.
\eeqa
The $R$ and $L$ sub-indices correspond to $P_R$ and $P_L$ which project
a Dirac higgsino onto the lower and upper Weyl fermion, respectively.
Thus, $C_{H^-R}$ is the coupling of the $H^-\overline{\widetilde
  N}P_R\widetilde C^+$ term, while $C_{H^-L}^*$ is the coupling of the
$H^-\overline{\widetilde N}P_L\widetilde C^+$ term.

To get some intuition about the expected size of the correction to the
MSSM annihilation cross section, we can estimate from the above
expressions that the relative size of the correction, $\kappa$:
\beq\label{eq:kap}
\kappa\sim 0.1 \left(\frac{5\text{ TeV}}{M/\lambda_1}\right)
\left(\frac{1-R_{\tilde\lambda}}{0.1}\right)^{1/2}.
\eeq
Thus, if there is a new physics threshold at around $5$ TeV, and the
higgsino component in the LSP is of order ten percent, then the
correction to the annihilation cross section is of order ten percent.
To make contact with the scalar spectrum, it is also useful to
represent Eq. (\ref{eq:kap}) in terms of $\epsilon_1$ and $\mu$,
\beq\label{eq:kapmu}
\kappa\sim0.6\left(\frac{m_W}{\mu}\right)\left(\frac{\epsilon_1}{0.1}\right)
\left(\frac{1-R_{\tilde\lambda}}{0.1}\right)^{1/2}.
\eeq
As the new physics threshold scales like $\sim\mu/\epsilon_1$, it is
difficult to envisage a phenomenologically acceptable scenario
exhibiting $\mu\sim m_W$ simultaneously with $\epsilon_1\sim0.1$.
Hence some suppression is to be expected from the combination of mass
and $\epsilon$ factors in (\ref{eq:kapmu}). This implies that the
BMSSM modification to the relevant MSSM annihilation processes is at
most of $\mathcal{O}(10\%)$ in generic cases.

\subsection{Two scalars}
We consider terms of the form
\beq
\overline{\tilde N_1}(y_{ab}^r+iy_{ab}^i\gamma^5)
\tilde N_1\phi_a\phi_b,
\eeq
where $y_{ab}$ has mass dimension of $-1$. Keeping only potential
$s$- and $p$-wave contributions, we obtain for the annihilation cross
section into $\phi_a\phi_b$:
\beq\label{sigann}
\sigma_{ab}v=\frac{\bar\beta_f c_{ab}^2 m_{\tilde N_1}^2}{4\pi
  S}\left[(y_{ab}^i)^2+\frac{v^2}{4}\left((y_{ab}^r)^2+\frac32
    (y_{ab}^i)^2\right)\right],
\eeq
where $m_{\tilde N_1}$ is the LSP mass,  $c_{ab}=2(1+\delta_{ab})$,
$S$ is the center-of-mass energy squared,
\beq
S\approx 4m_{\tilde N_1}^2(1+v^2/4),
\eeq
and
\beq
\bar\beta_f\equiv\frac{2|\vec p_a|}{\sqrt{S}}\approx
\sqrt{1+\frac{(m_a-m_b)^2}{16m_{\tilde
      N_1}^4}-\frac{m_a^2+m_b^2}{2m_{\tilde N_1}^2}}.
\eeq

In terms of the Lagrangian parameters, we have
\beqa
y_{ab}^r&=&-\frac{1}{2M}{\cal R}e\left(\lambda_1^* Y_{ab}\right),\ \ \
y_{ab}^i=-\frac{1}{2M}{\cal I}m\left(\lambda_1^* Y_{ab}\right ),\ \ \
\phi_a\phi_b=hh,HH,AA,hH,H^+H^-\no\\
y_{ab}^r&=&\frac{1}{2M}{\cal I}m\left(\lambda_1^* Y_{ab}\right),\ \ \
y_{ab}^i=-\frac{1}{2M}{\cal R}e\left(\lambda_1^* Y_{ab}\right),\ \ \
\phi_a\phi_b=hA,HA.
\eeqa
The dimensionless $Y_{ab}$ couplings are given by
\beqa\label{eq:Yhh}
Y_{hh}&=&s^2_\alpha Z_{14}^2+c_\alpha^2 Z_{13}^2-2s_{2\alpha} Z_{14}Z_{13},\no\\
Y_{HH}&=&c^2_\alpha Z_{14}^2+s_\alpha^2 Z_{13}^2+2s_{2\alpha} Z_{14}Z_{13},\no\\
Y_{AA}&=&-s^2_\beta Z_{14}^2-c_\beta^2 Z_{13}^2-2s_{2\beta} Z_{14}Z_{13},\no\\
Y_{hH}&=&s_{2\alpha}(-Z_{14}^2+Z_{13}^2)+4c_{2\alpha} Z_{14}Z_{13},\no\\
Y_{hA}&=&2s_\alpha s_\beta Z_{14}^2-2c_\alpha c_\beta
Z_{13}^2+4s_{(\alpha-\beta)} Z_{14}Z_{13},\no\\
Y_{HA}&=&-2c_\alpha s_\beta Z_{14}^2-2s_\alpha c_\beta Z_{13}^2-4c_{(\alpha-\beta)} Z_{14}Z_{13},\no\\
Y_{H^+H^-}&=&-2s_{2\beta} Z_{13}Z_{14}.
\eeqa

Co-annihilation proceeds via terms of the form
\beq
\overline{\tilde N_1}(y_{ab}^e+y_{ab}^o\gamma^5)\tilde C^+_1,
\eeq
with $y^e,y^o$ complex. The cross-section is given by an expression
similar to Eq. (\ref{sigann}), provided that the neutralino-chargino
mass difference is neglected, taking $c_{ab}=1$, and making the
substitution $(y^r)^2\to|y^e|^2, \ \ (y^i)^2\to|y^o|^2$. We now have
\beqa
y_{H^-\phi_b}^{e,o}&=&-\frac{1}{2M}\left(\lambda_1^*Y_{H^-\phi_b R}\pm
\lambda_1 Y^*_{H^-\phi_b L}\right),\ \ \ \phi_b=h,H,\no\\
y_{H^-A}^{e,o}&=&-\frac{1}{2M}\left(\lambda_1^*Y_{H^-A R}\mp \lambda_1 Y^*_{H^-A L}\right), \eeqa
\beqa\label{eq:YH-h}
Y_{H^- h R}&=&\sqrt{2}c_\beta U_{12}(s_\alpha Z_{14}-c_\alpha Z_{13}),\no\\
Y_{H^- h L}&=&\sqrt{2}s_\beta V_{12}(s_\alpha Z_{14}-c_\alpha Z_{13}),\no\\
Y_{H^- H R}&=&-\sqrt{2}c_\beta U_{12}(c_\alpha Z_{14}+s_\alpha Z_{13}),\no\\
Y_{H^- H L}&=&-\sqrt{2}s_\beta V_{12}(c_\alpha Z_{14}+s_\alpha Z_{13}),\no\\
Y_{H^- A R}&=&\sqrt{2}c_\beta U_{12}(s_\beta Z_{14}+c_\beta Z_{13}),\no\\
Y_{H^- A L}&=&\sqrt{2}s_\beta V_{12}(s_\beta Z_{14}+c_\beta Z_{13}).
\eeqa

To get some intuition about the expected size of the correction to the
MSSM annihilation cross section, let us consider the case of a
higgsino LSP, $R_{\tilde\lambda}\approx0$, wherein the couplings
(\ref{eq:Yhh}) and (\ref{eq:YH-h}) are unsuppressed. In the MSSM,
annihilation into (mostly transverse) gauge bosons gives
$\langle\sigma v\rangle\sim \frac{g^4}{85\pi\mu^2}$
\cite{ArkaniHamed:2006mb}. Regarding the BMSSM contribution
(\ref{sigann}) to annihilation into light Higgs boson pairs, we obtain
$\langle\sigma v\rangle\sim\frac{\lambda_1^2}{24\pi M^2}$. This
estimate holds when there is no CP violation and the leading
contribution is $p$-wave. Writing the modified cross section as
$\langle\sigma v\rangle=\langle\sigma v\rangle_0(1+\delta_\epsilon)$,
we find
$\delta_\epsilon\sim0.25\left(\frac{\epsilon_1}{0.1}\right)^2$. Note,
however, that in the relevant scenario co-annihilations are important
and so further numerical study is required to assess the full impact
of the BMSSM. Below, we proceed to perform this study.

\section{The dark matter relic density}
\label{sec:dmrel}
As deduced from the WMAP satellite measurement of the temperature
anisotropies in the Cosmic Microwave Background, cold dark matter
makes up approximately $23\%$ of the energy of the Universe
\cite{Dunkley:2008ie}. The DM cosmological density is precisely
measured to be
\begin{equation}
\Omega_{DM}\,h^2=0.101\pm 0.062
\end{equation}
at $68\%$ CL. The accuracy is expected to be improved to the percent
level by future measurements at Planck satellite \cite{Bouchet:2007zz}.

We calculate the dark matter relic density in the presence of
the $\epsilon_1$ couplings using a modified version of the code
MicrOMEGAs \cite{micromegas}, where we implemented the BMSSM
Higgs-Higgs-higgsino-higgsino couplings of Eq. (\ref{eq:hhHH}).
The leading $\epsilon_{1,2}$-induced corrections to the MSSM Higgs
spectrum, Eq. \eqref{eq:delm}, were implemented using the code
SuSpect \cite{Djouadi:2002ze}.

The BMSSM framework, if relevant to the little hierarchy problem that
arises from the lower bound on the Higgs mass, assumes a new physics
scale at a few TeV. Since the new degrees of freedom at this scale are
not specified, the effect of the new threshold on the running of parameters
from a much higher scale cannot be rigorously taken into account. It
therefore only makes sense to study the BMSSM effects in a framework
specified at low energy. In order to demonstrate some of the most
interesting consequences of the BMSSM operators for dark matter, we will
employ two such sets of parameters: a model where all sfermion masses are
correlated, and a model where the only light sfermions are the stops.
The first model demonstrates how the so-called bulk region is re-opened,
even for correlated stop and slepton masses. The second model incorporates
the interesting process of stop co-annihilation. For both models we focus
our attention mainly on regions where the stops are light, since the main
motivation for the BMSSM operators is to avoid a heavy stop (which is the
cause of the little hierarchy problem) and, furthermore, this is the region
that is relevant to BMSSM baryogenesis. Previous analysis in the context
of the MSSM with a light stop were done in \cite{DFL,Balazs:2004bu,Balazs:2004ae}.

\subsection{Correlated stop-slepton masses}
The most natural dark matter scenario within the MSSM framework could
have been that of a light bino, annihilating to the standard model
leptons via light slepton exchange. This scenario is known as the
``bulk region'' of the MSSM. However, in some of the most intensively
studied MSSM scenarios, such as the mSUGRA \cite{mSUGRA} or cMSSM
frameworks, the part of the bulk region that is allowed became smaller
and smaller as the experimental lower bound on the Higgs mass became
stronger. The generic reason for this is that a stronger lower bound
on the Higgs mass requires a heavier stop which, in these frameworks,
further implies heavy sleptons. One way to re-open the bulk region is
to assume a framework where the stop and the slepton masses are not
correlated. The BMSSM, however, re-opens the bulk region in a
different way: the stop is not required to be heavy anymore.

In order to understand these implications of the BMSSM framework and,
in particular, in order to allow for a simple comparison with
mSUGRA-like models, we investigate the following framework. The MSSM
parameters that we use are those that would have corresponded to an
mSUGRA model specified by the five parameters
\beq\label{eq:msugra}
\tan\beta,\ m_{1/2},\ m_0,\ A_0,\ {\rm sign}(\mu).
\eeq
Thus, the correlations between the low energy MSSM parameters are
the same as those that would hold in an mSUGRA framework.
In other words, our low energy parameters are expressed in terms
of the parameters in (\ref{eq:msugra}) approximately as follows
\cite{Drees:1992am}:
\beqa
m^2_{\tilde q}&\approx&m_0^2+6\,m_{1/2}^2,\no\\
m^2_{\tilde\ell_L}&\approx&m_0^2+0.5\,m_{1/2}^2,\no\\
m^2_{\tilde\ell_R}&\approx&m_0^2+0.15\,m_{1/2}^2,\no\\
M_1&\approx&0.4\,m_{1/2},\no\\
M_2&\approx&0.8\,m_{1/2},\no\\
M_3&\approx&3\,m_{1/2}.
\eeqa
The values of $\mu^2$ and $m_A^2$ depend on the soft breaking
terms and on the electroweak breaking parameters in the standard way.
Let us emphasize again that one should {\it not}
think about this set of parameters as coming from an extended mSUGRA
model, since the effects of the BMSSM physics at the few TeV scale on
the running are not (and cannot) be taken into account. In addition,
we have two extra BMSSM parameters: $\epsilon_1$ and $\epsilon_2$. We
focus essentially on the effects of $\epsilon_1$.

\begin{figure}[ht!]
\begin{center}
\vspace{-1.2cm}\hspace{-2.5cm}
\includegraphics[width=10.7cm]{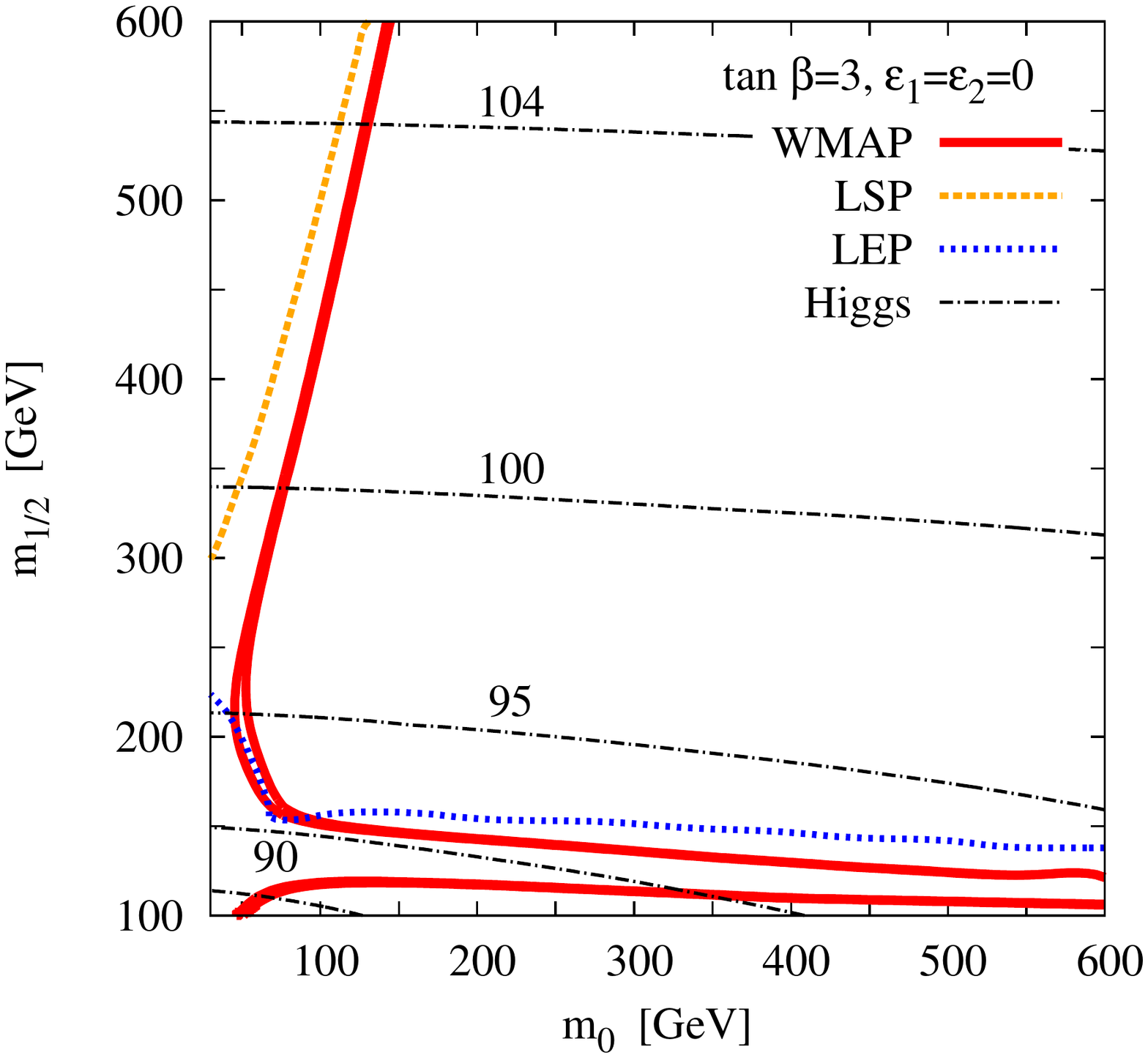}\hspace{-3.3cm}
\includegraphics[width=10.7cm]{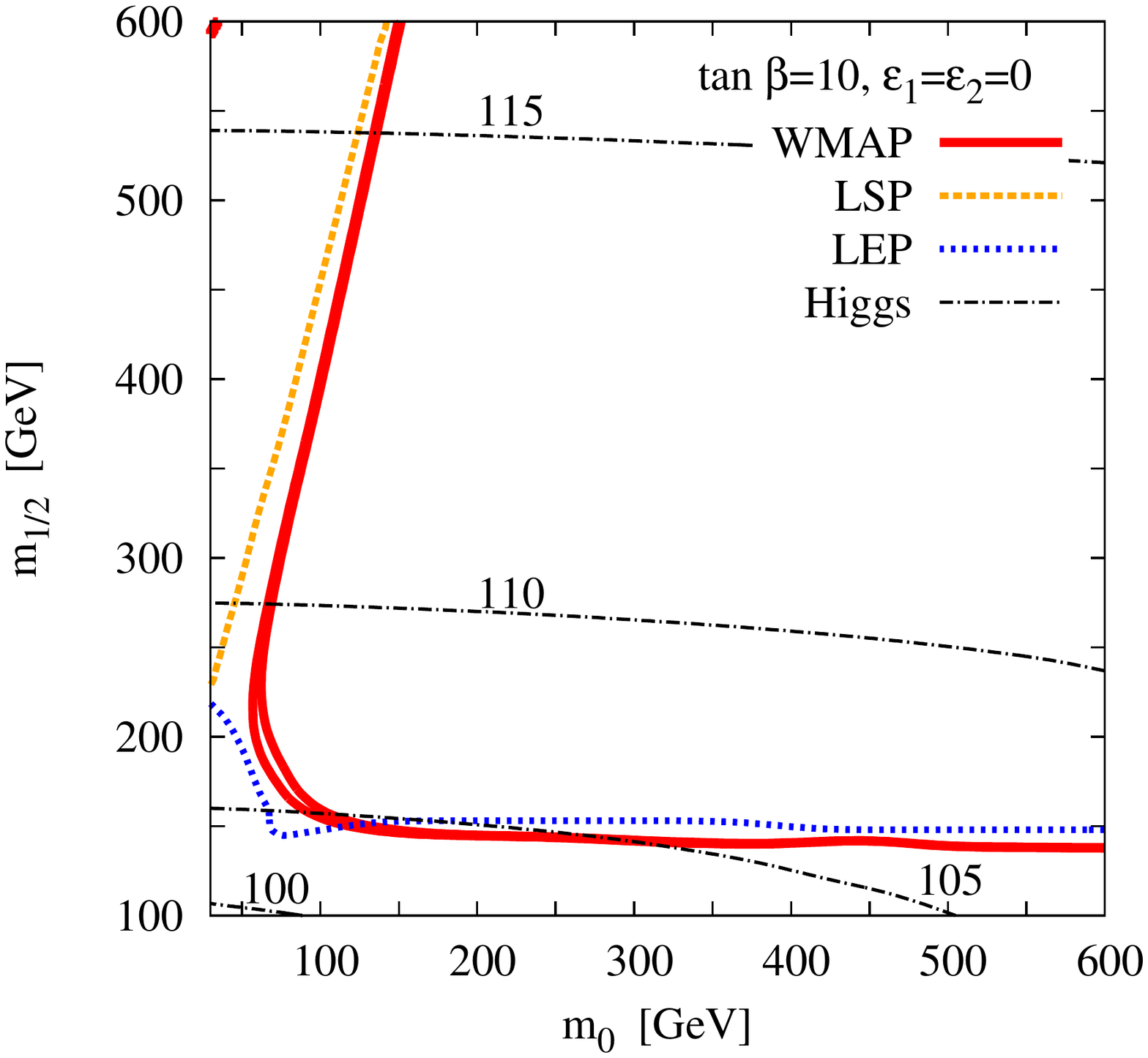}\\
\vspace{-1.7cm}\hspace{-2.5cm}
\includegraphics[width=10.7cm]{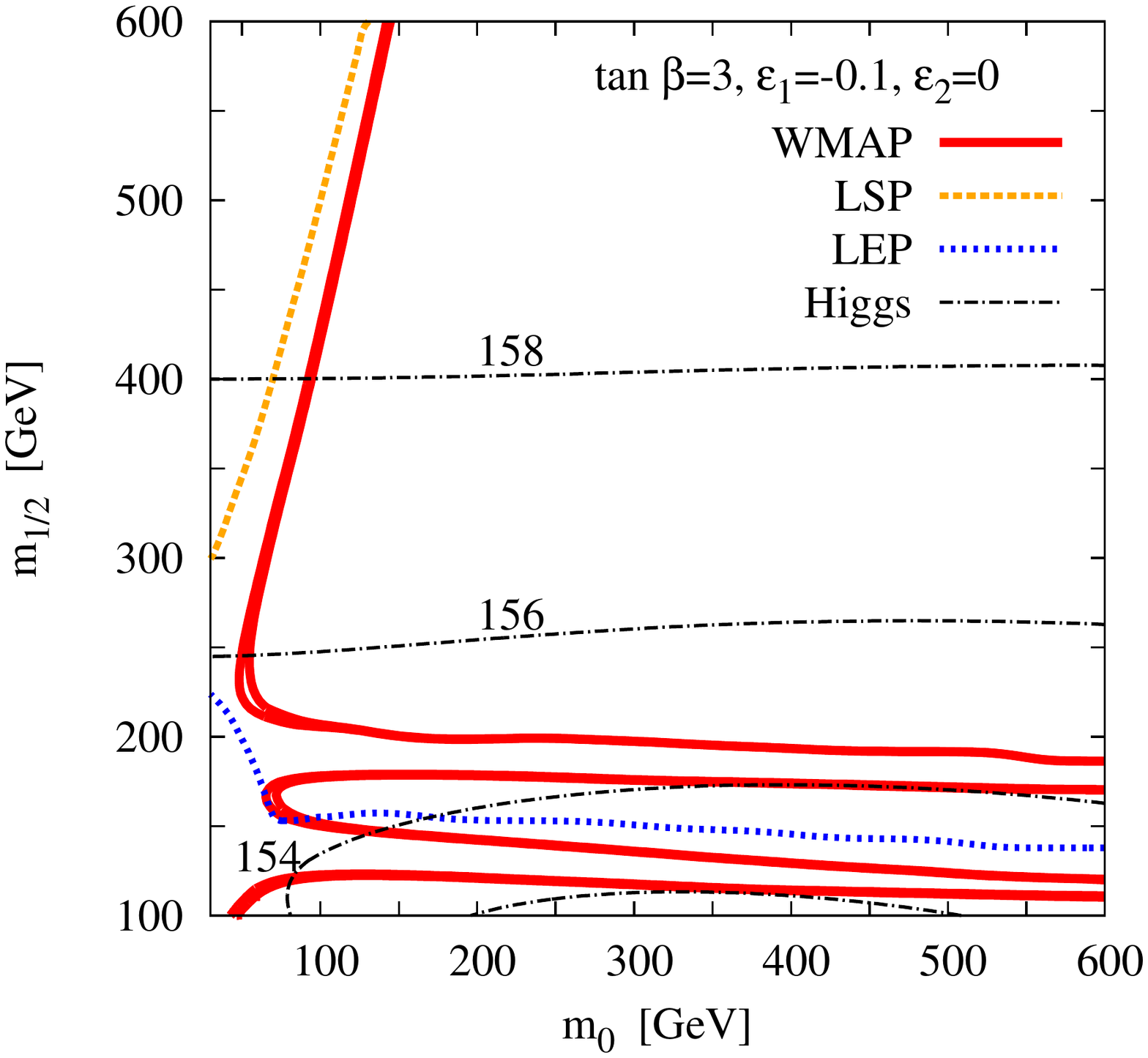}\hspace{-3.3cm}
\includegraphics[width=10.7cm]{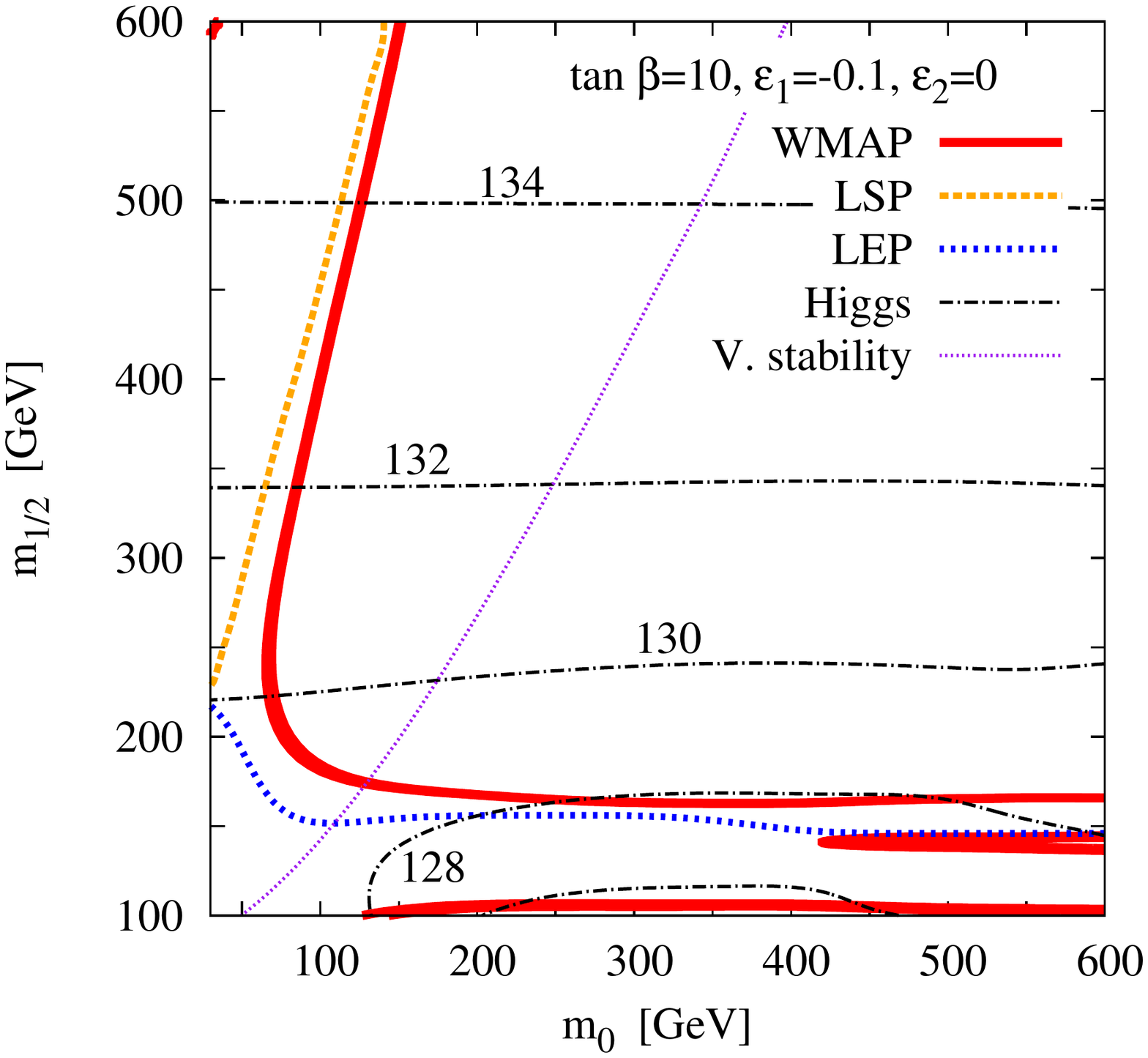}\\
\vspace{-1.7cm}\hspace{-2.5cm}
\includegraphics[width=10.8cm]{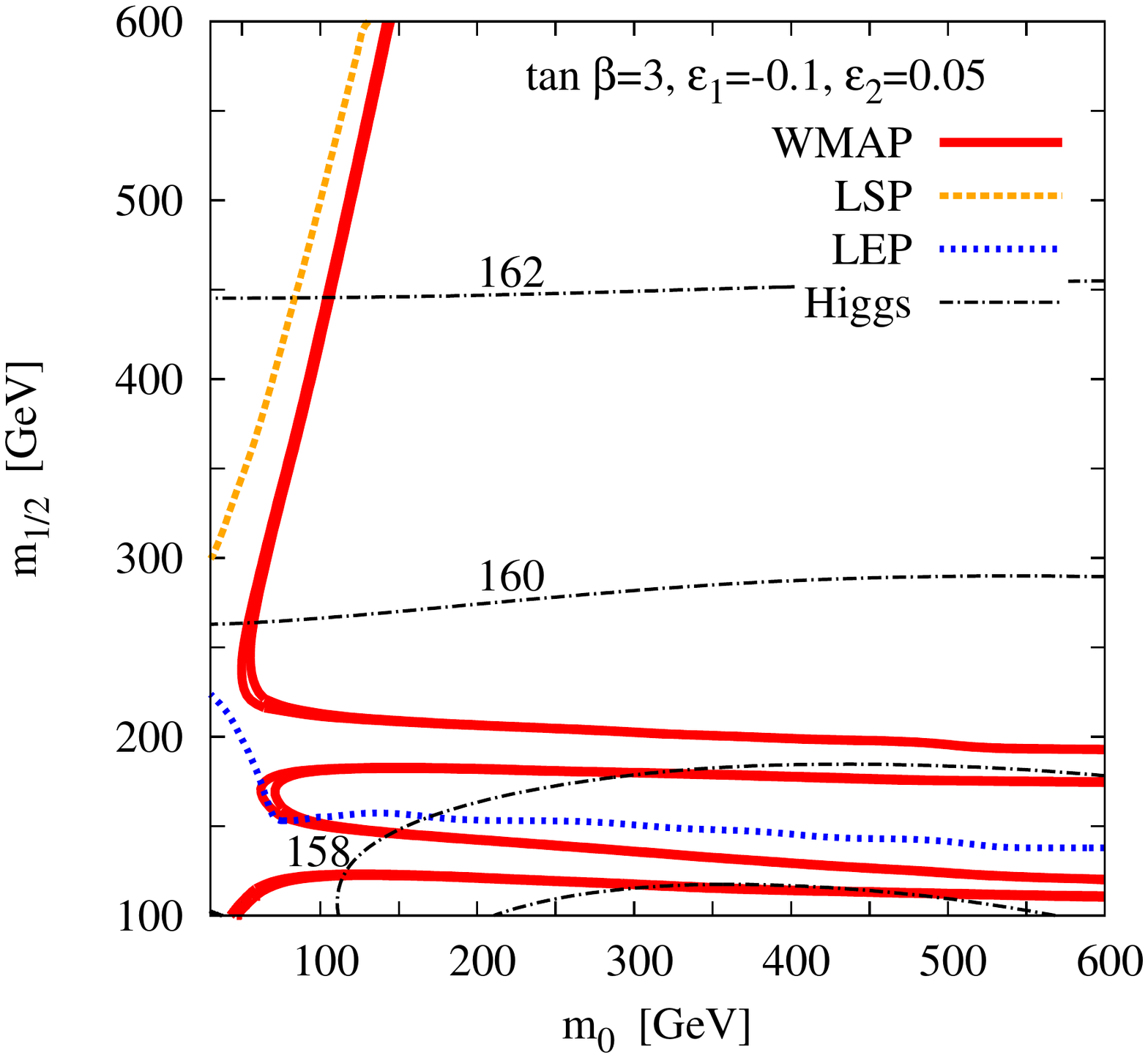}\hspace{-3.3cm}
\includegraphics[width=10.8cm]{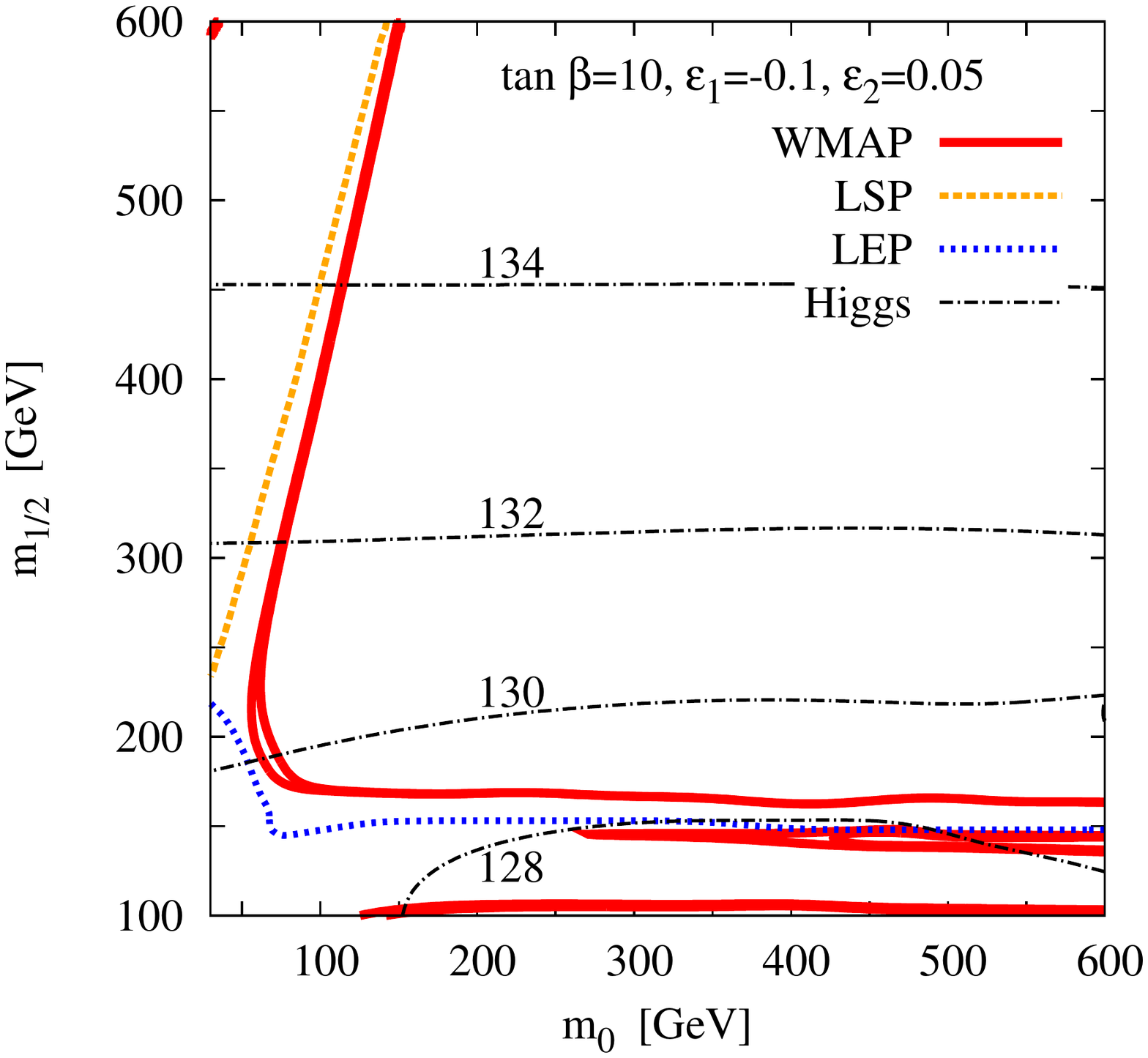}
\end{center}
\vspace{-1.9cm}
\caption{Regions in the $[m_0,\,m_{1/2}]$ plane in which the WMAP
  constraint is fulfilled (between the solid red lines). We use
  $A_0=0$ GeV, $\mu>0$, and values of $\tan\beta$, $\epsilon_1$ and
$\epsilon_2$ as indicated in the various panels. The region below the
  dotted blue curve is excluded by the null searches for charginos and
  sleptons at LEP. The area to the left and above the orange curve is
  excluded because the stau is the LSP. The dash-dotted black curves
  are contour lines for $m_h$ with values in GeV as indicated. The
  dotted purple line in the middle right panel denotes the appearance
  of a remote vacuum in the scalar potential. Above and to the left of
  this line, the electroweak vacuum is metastable.}
\label{dm1}
\end{figure}

In practice, we make discrete choices of $\tan\beta$, $A_0$,
sign$(\mu)$ and $\epsilon_1$, and scan over $m_0$ and $m_{1/2}$. We
focus our attention on moderate values of $m_{1/2}$ and $m_0$ because
we are mainly interested in light sfermions and the bulk region. Fig.
\ref{dm1} displays the area, in the $[m_0,\,m_{1/2}]$ plane, where the
WMAP constraint is satisfied (between the solid red lines). We use
$A_0=0$ GeV, $\mu>0$, and values of $\tan\beta$, $\epsilon_1$ and
$\epsilon_2$ as indicated in the various plots. The region below the
dotted blue curve is excluded by the null searches for charginos and
sleptons at LEP.  The area to the left and above the orange curve is
excluded because the stau is the LSP. The dotted black curves are
contour lines for $m_h$.

We would like to emphasize several points regarding the effects that
are demonstrated by Fig. \ref{dm1}:
\begin{enumerate}
\item The most significant effect of the BMSSM operators is their
  impact on the Higgs boson mass. Within the MSSM with mSUGRA-like
  correlations, the bound on the Higgs mass strongly constrains
  $m_{1/2}$. In contrast, in the presence of $\epsilon_1={\cal
    O}(-0.05)$, the full region for which the correct value of the
  relic abundance is obtained is allowed.
\item
Another significant effect is the appearance (for negative enough
$\epsilon_1$ values) of a new region fulfilling the DM constraint.
This is the `$h$-pole' region in which $m_h\sim 2\,m_{\chi_1^0}$,
and the $s$-channel Higgs exchange is nearly resonant, allowing the
neutralinos to annihilate efficiently \cite{Djouadi:2005dz}.
\item For the case discussed in this subsection, the co-annihilation
  region with the stops is not present as the mass difference with the
  LSP is always too large.
\item In the $m_0$ region that we are considering here, the impact of
  the BMSSM operators on the mass of the neutralino LSP and on the
  region fulfilling the WMAP constraint is rather limited. The reason
  is that in the bulk region the LSP is mostly bino-like, while the
  BMSSM operators affect the higgsino parameters.
\end{enumerate}

The BMSSM operators may destabilize the scalar potential. If
$4|\epsilon_1|>\epsilon_2$, the effective quartic coupling along one
of the D-flat directions is negative, causing a remote vacuum to form
in the presence of which the electroweak vacuum could become
metastable. When considering values of $\epsilon_1\gsim-0.1$, vacuum
stability is ensured provided that the following condition is
fulfilled \cite{Blum:2009na}:
\beq\label{eq:VstabEW}
\frac{m_A^2}{|\mu|^2}\geq\frac{2}{1+\sin2\beta}
\left(1+\frac{\epsilon_2}{4\epsilon_1}\right)^2.
\eeq
The stability criterion as written here applies when $m_A\gsim3m_Z$.
The fact that this criterion involves electroweak scale parameters,
and only ratios of non-renormalizable operators, is reminiscent of the
supersymmetric origins of $\epsilon_1$. In regions of the parameter
space where Eq.  (\ref{eq:VstabEW}) is violated, the electroweak
vacuum is metastable.  For large $\tan\beta$, this occurs throughout
an important fraction of the $[m_0,\,m_{1/2}]$ plane, depicted in the
middle right panel of Fig. \ref{dm1} by the area above and to the left
of the dotted purple line.

It is important to stress that Eq.~(\ref{eq:VstabEW}) represents an
analytical tree level approximation and as such, in the current
framework where quantum corrections are sizeable, it is
conservative. In order to determine whether the lifetime of the
electroweak vacuum is long enough, the rate of quantum tunneling into
the remote vacuum should be compared to the Hubble rate.
For illustration, consider the two parameter points,
$P_1: \ [m_0=200\ {\rm GeV},m_{1/2}=150\ {\rm GeV}]$ and
$P_2: \ [m_0=100\ {\rm GeV},m_{1/2}=250\ {\rm GeV}]$, with
$\tan\beta=10,\epsilon_1=-0.1,\epsilon_2=0$, as in the middle right
panel of Fig.~\ref{dm1}. Point $P_1$ corresponds to a stable
electroweak vacuum configuration, in agreement with the
criterion~(\ref{eq:VstabEW}). Point $P_2$, on the other hand, violates
Eq.~(\ref{eq:VstabEW}) by $\sim15\%$. However, $P_2$ involves
significant quantum corrections due to stops ($m_{\tilde t_1}=415$ GeV, $m_{\tilde t_2}=590$
GeV, $A_t=-480$ GeV). Calculating the tunneling action, we find that
$P_2$ is in fact long-lived enough to provide an acceptable model
point. Similar calculations reveal that the rest of the middle right
panel is also long-lived enough (at least marginally), and so
Eq.~(\ref{eq:VstabEW}) can not be used to exclude regions in the
mSUGRA-like parameter space considered in this section. We will return
to the issue of vacuum stability in Section~\ref{ssec:lshs}. There,
quantum corrections due to stops will be held moderate and fixed,
resulting with precise application of Eq.~(\ref{eq:VstabEW}) to
exclude significant portions of the parameter space.

As concerns precision electroweak data and low energy processes, it is
important to realize that the new physics that generates the
non-renormalizable operators can directly modify the constraints that
come from these measurements. Ignoring this point, it is still
possible to identify regions in the parameter space favored by the
WMAP data which satisfy all such low energy constraints. The relevance
of the BMSSM lies in the fact that constraints involving the Higgs are
decoupled from constraints involving the stop sector. In particular,
stops that are neither heavy nor mixed are acceptable, as demonstrated
by our choice of $A_0=0$ throughout the current section. In contrast,
in the case of the MSSM, satisfying the Higgs mass bound as well as
electroweak precision data necessitates large values for $A_0$
\cite{Belyaev:2007xa}.

\subsection{Light stops, heavy sleptons}
\label{ssec:lshs}

The aim of this section is to further expose various implications of
the BMSSM for the DM relic abundance, putting special emphasis on
parameter regions compatible with a strong first order electroweak
phase transition, as required for baryogenesis. In particular, we are
interested in the scenario of light, unmixed stops. As mentioned
above, LSP co-annihilation with stops is not a viable possibility if
the low energy soft supersymmetry breaking parameters obey relations
similar to those that would follow from mSUGRA-like theory, unless
stop mixing is very large. To explore this possibility in the BMSSM,
we employ a set of low energy parameters that is different from the
previous subsection. Explicitly, in addition to the BMSSM $\epsilon_i$
parameters, we consider the following set of parameters: \beq M_2,\
\mu,\ \tan\beta,\ X_t,\ m_U,\ m_Q, \ m_{\tilde f},\ m_A, \eeq where
$m_{\tilde f}$ is a common mass for the sleptons, the first and second
generation squarks, and $\tilde b_R$. We further use
$M_1=\frac53\,\tan^2\theta_W\,M_2\sim\frac12\,M_2$.

To demonstrate our main points, we fix the values of all but two
parameters as follows: $\epsilon_1=0$ or $-0.1$, $\epsilon_2=0$ or
$+0.05$, $\tan\beta=3$ or $10$, $X_t=0$, $m_U=210$ GeV, $m_Q=400$ GeV,
$m_{\tilde f}=m_A=500$ GeV. This scenario gives rise to relatively
light stops:
\beq
m_{\tilde t_1}\lesssim 150\ {\rm GeV},\ \ \ \ 370\
{\rm GeV}\lesssim m_{\tilde t_2}\lesssim 400\ {\rm GeV}.
\eeq
We scan over the remaining two parameters, $M_2$ and $\mu$.

In the prescribed framework, one can identify four regions in which
the WMAP constraint is fulfilled:
\begin{itemize}
\item[--] The `$Z$-pole' region in which the LSP is very light,
  $m_{\chi_1^0}\sim\frac12 M_Z\sim 45$ GeV, and the $s$-channel $Z$
  exchange is nearly resonant. This region is not ruled out only in
  scenarios where the mass splitting between $M_1$ and $M_2$ at the
  electroweak scale is very large.
\item[--] The `$h$-pole' region in which the LSP is rather light,
  $m_{\chi_1^0}\sim\frac12 M_h$, and the $s$-channel $h$ exchange is
  nearly resonant, allowing the neutralinos to annihilate efficiently
  \cite{Djouadi:2005dz}.
\item[--] The `mixed region' in which the LSP is a higgsino--bino
  mixture \cite{ArkaniHamed:2006mb}, $M_2 \sim 2\mu$, which enhances
  (but not too much) its annihilation cross-sections into final states
  containing gauge and/or Higgs bosons: $\chi_1^0 \chi_1^0 \to W^+
  W^-$, $ZZ$, $Zh$ and $hh$.
\item[--] The `stop co-annihilation' region, in which the LSP is
  almost degenerate in mass with the lightest stop ($\tilde t_1$).
  Such a scenario leads to an enhanced annihilation of sparticles
  since the $\chi_1^0-\tilde t_1$ co-annihilation cross-section
  \cite{Griest:1990kh,Ellis:2001nx} is much larger than that of the
  LSP.
\end{itemize}

\begin{figure}[ht!]
\begin{center}
\vspace{-1.1cm}\hspace{-2.3cm}
\includegraphics[width=10.9cm]{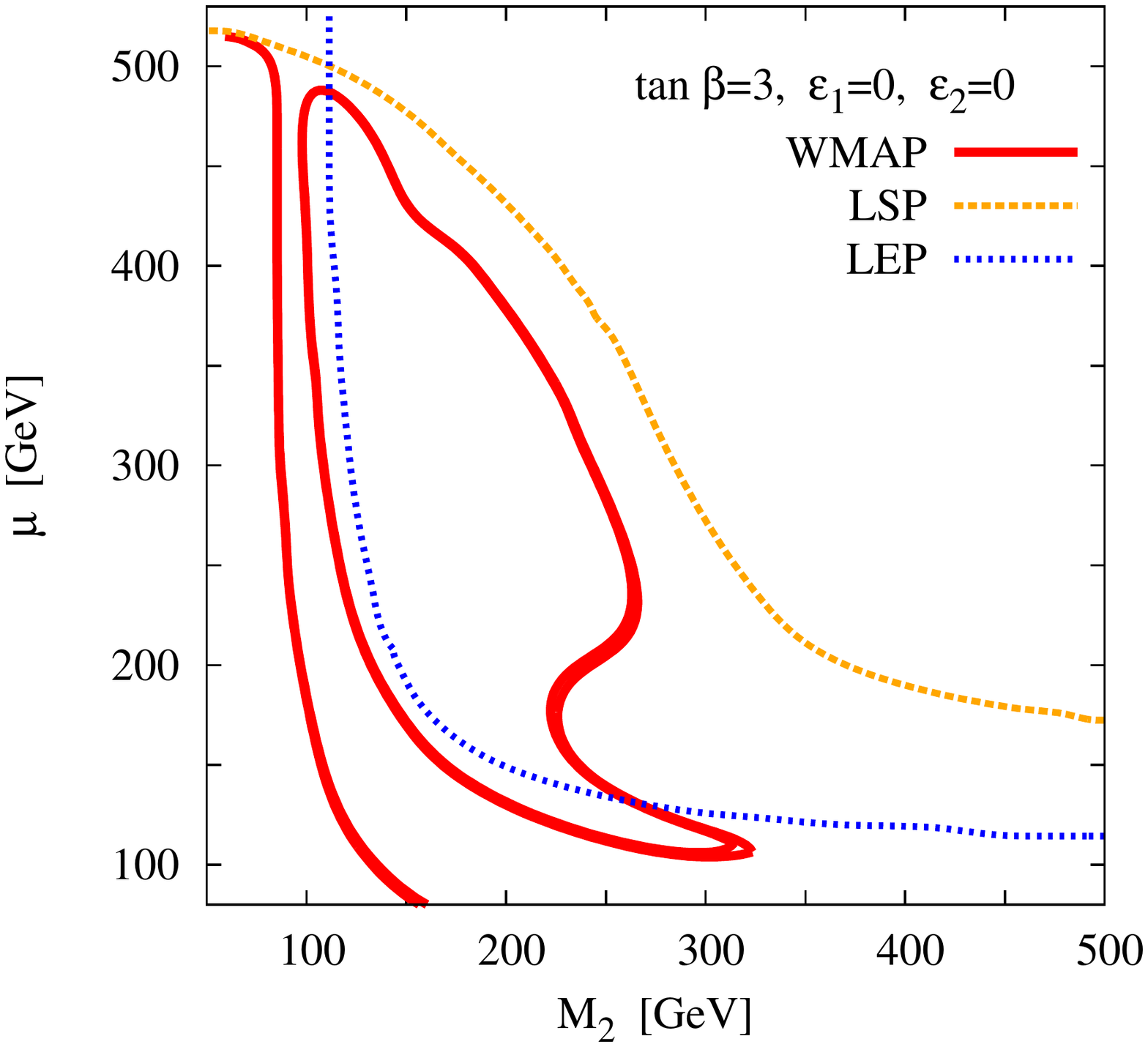} \hspace{-3.9cm}
\includegraphics[width=10.9cm]{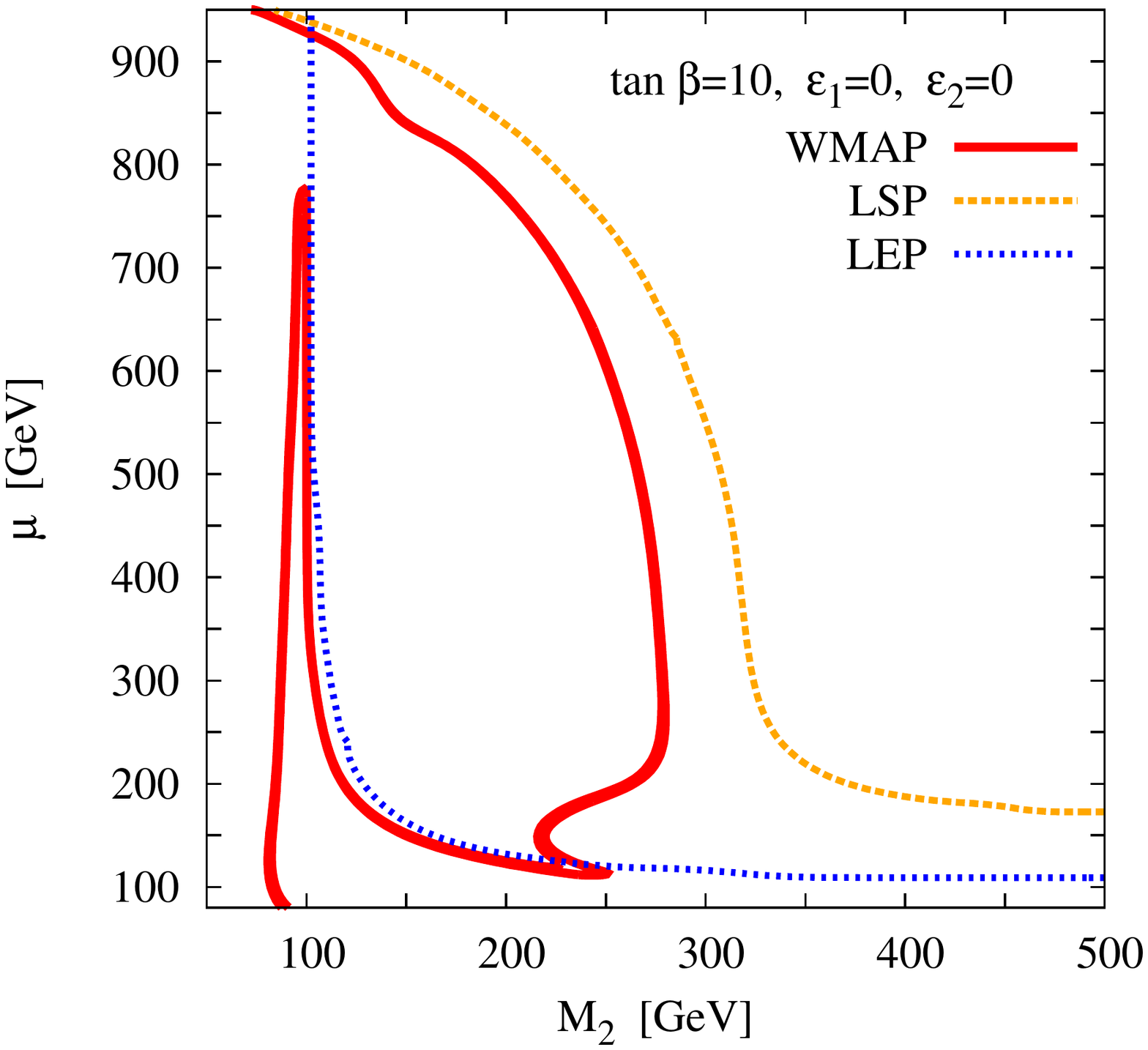}\\
\vspace{-1.6cm}\hspace{-2.3cm}
\includegraphics[width=10.9cm]{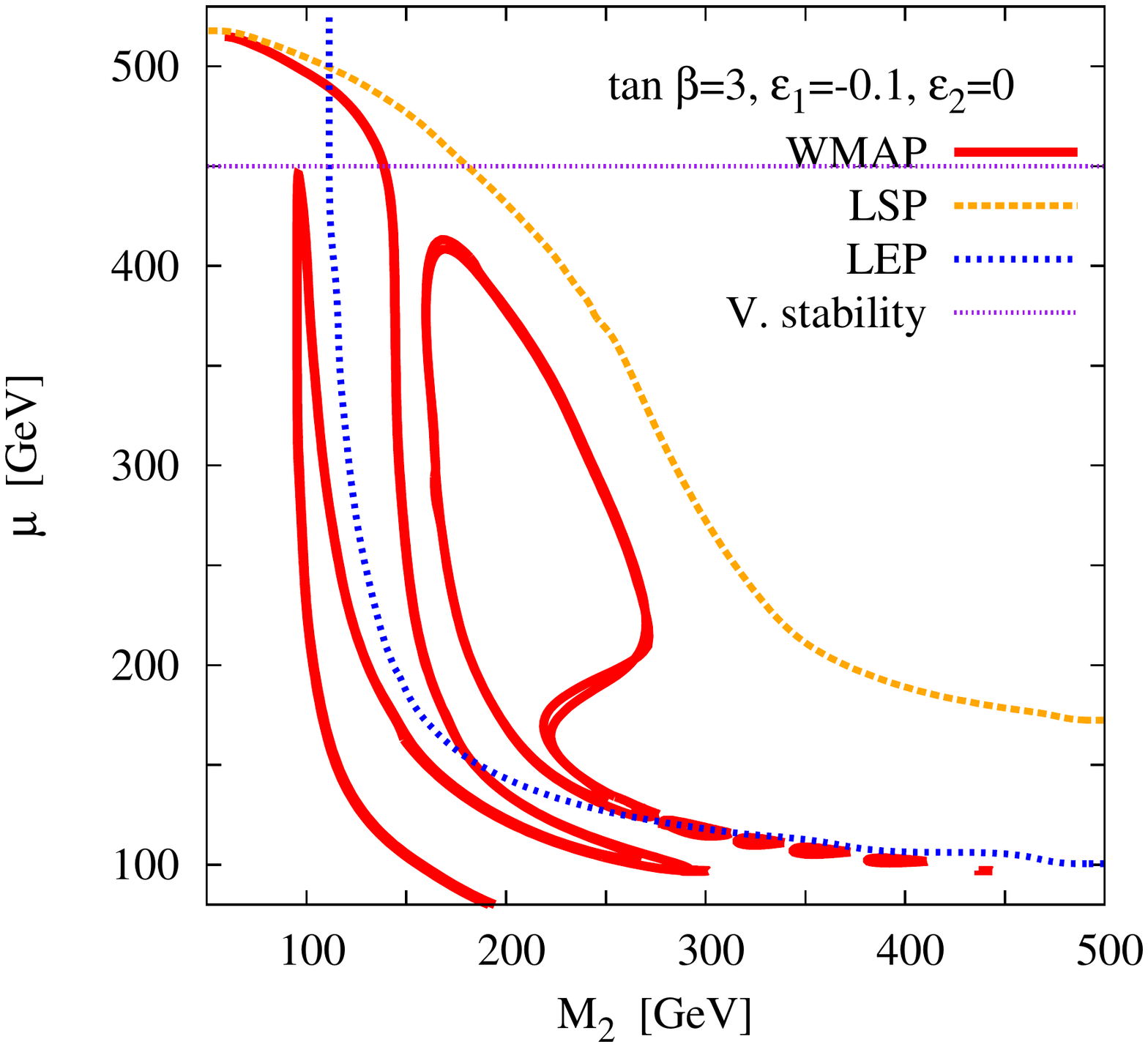}\hspace{-3.9cm}
\includegraphics[width=10.9cm]{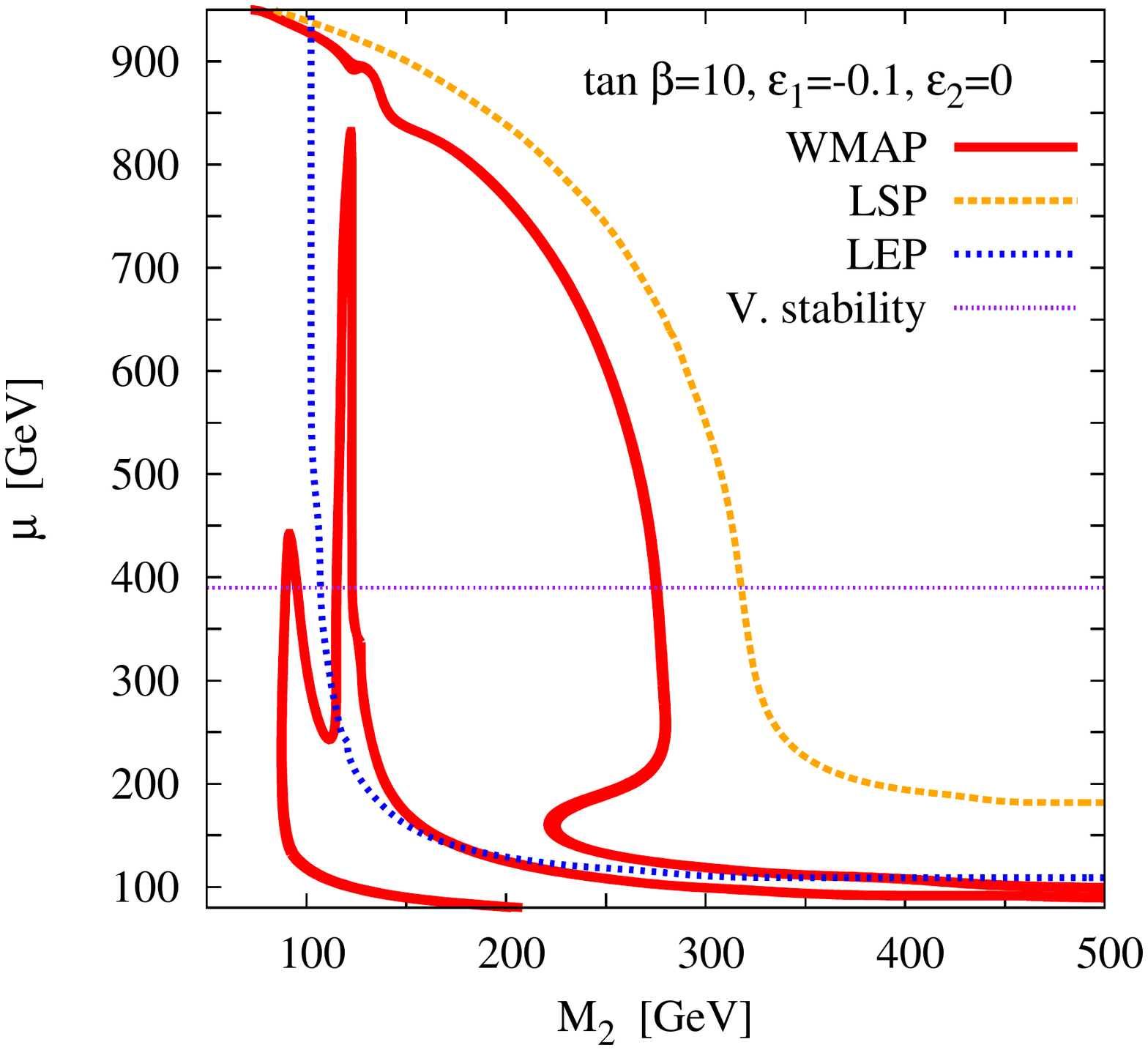}\\
\vspace{-1.6cm}\hspace{-2.3cm}
\includegraphics[width=10.9cm]{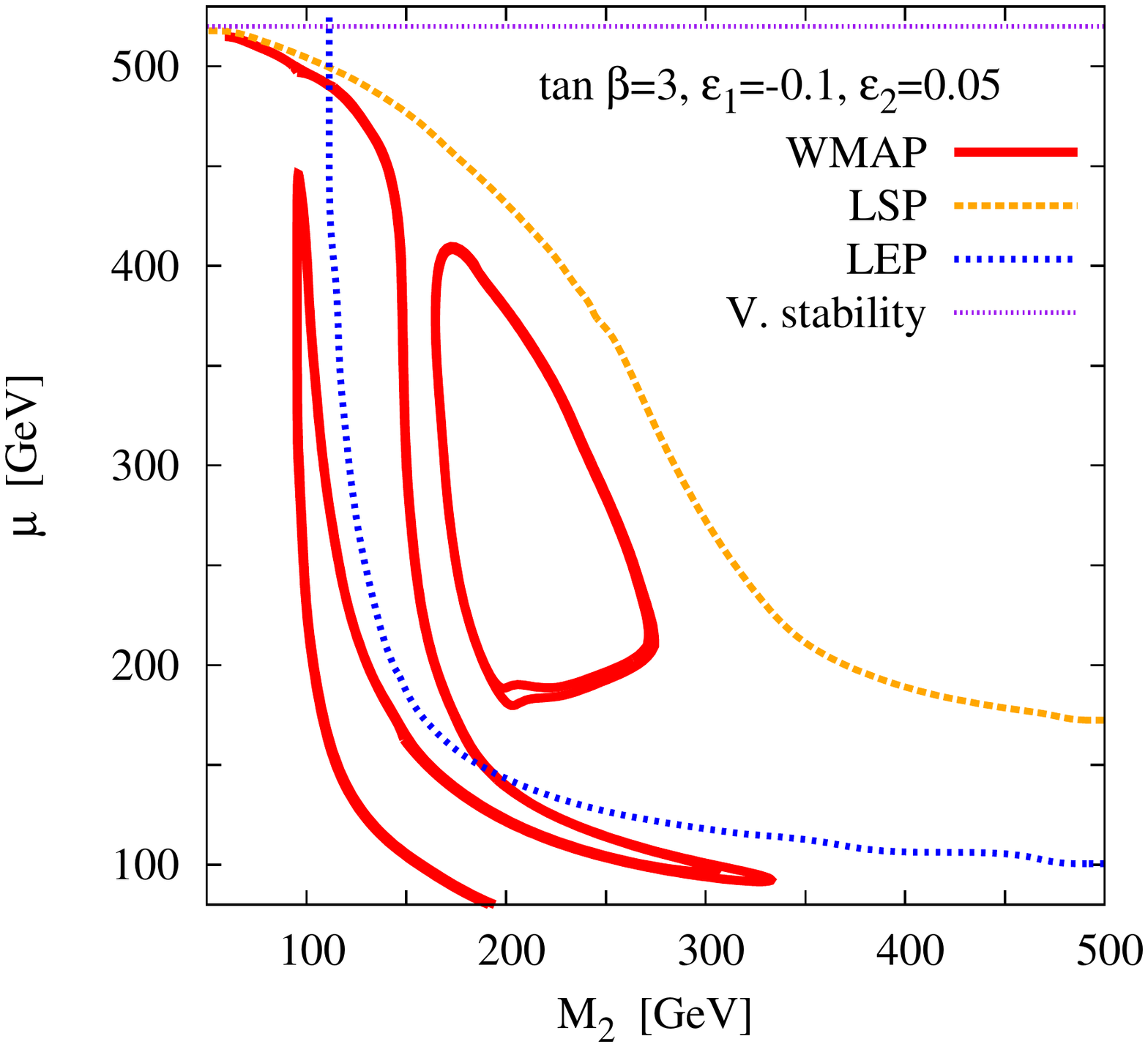}\hspace{-3.9cm}
\includegraphics[width=10.9cm]{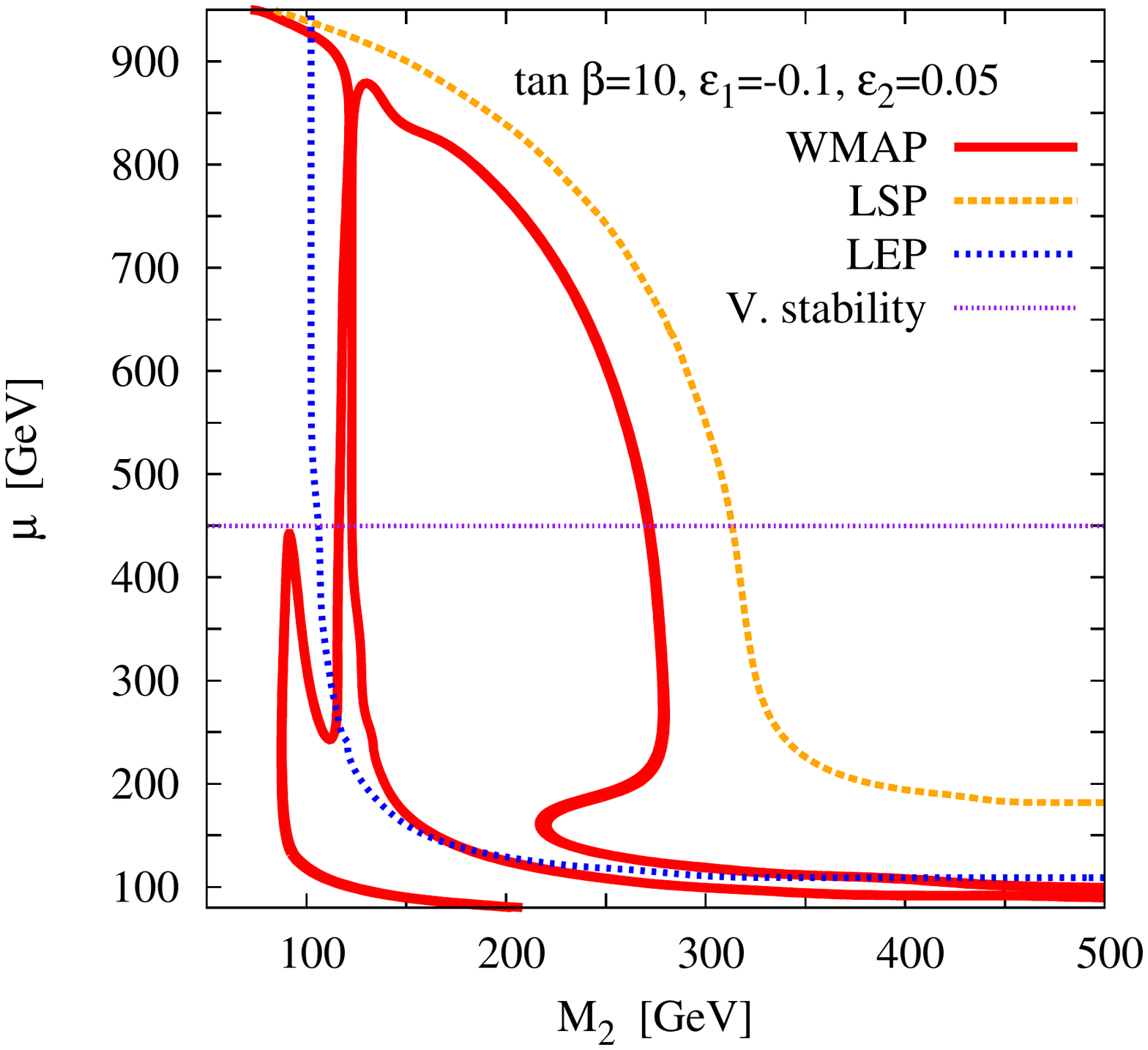}
\end{center}
\vspace{-1.8cm} \caption{Regions of the $[M_2,\,\mu]$ plane in which
  the WMAP constraint is fulfilled (between the solid red lines). We
  use the parameter set described in the text, and values of
  $\tan\beta$, $\epsilon_1$ and $\epsilon_2$ as indicated in the
  various panels. The region above and to the right of the orange
  dashed curve corresponds to a stop LSP. The region to the left and
  below the blue dotted curve is excluded by direct chargino
  searches. Most of the region above the horizontal dashed line is
  excluded by vacuum instability.}
\label{dm4}
\end{figure}

Fig. \ref{dm4} displays the areas, in the $[M_2,\,\mu]$ plane, in
which the WMAP constraint is satisfied (between the solid red lines).
The region of large $\mu$ and $M_2$, to the right of the orange dashed
line, is excluded since the lightest stop becomes the LSP. The region
of small $\mu$ and/or $M_2$, below and to the left of the blue dotted
line, is excluded by the null search for charginos at LEP2
\cite{Amsler:2008zzb}.

Let us first consider the two upper panels, where
$\epsilon_1=\epsilon_2=0$.  The various regions described above that
are consistent with the DM constraints can be identified in this
figure.  The region at $M_2 \sim m_Z\sim m_h$ corresponds to the
$s$-channel exchange of an almost on-shell Higgs or $Z$ boson. Note
that when $2\,m_{\chi_1^0}$ is too close to the Higgs or $Z$ mass
pole, the LSP annihilation is too efficient and leads to a much too
small $\Omega_{\rm DM}\,h^2$. In any case, for the Higgs mass values
obtained here, $m_h \sim 85(98)$ GeV for $\tan\beta=3(10)$, this
region is already excluded by the negative searches for chargino pairs
at LEP2.

The region close to $M_2\sim\mu\sim 200$ GeV corresponds to the LSP
being a bino--higgsino mixture with sizeable couplings to $W$, $Z$ and
Higgs bosons, allowing for reasonably large rates for neutralino
annihilation into $\chi_1^0 \chi_1^0 \to W^+W^-$, $ZZ$, $hZ$ and $hh$
final states. Above and below the band, the LSP couplings to the
various final states are either too strong or too weak to generate the
relevant relic density.

Finally, for larger $\mu$ values, in the region close to the orange
curve, the mass of the lightest neutralino approaches the mass of the
lightest stop leading to an enhanced co-annihilation cross-section:
$\chi_1^0\,\tilde t_1\to W^+\,b$, $g\,t$ ($\sim 90\%$). Also, to a
lesser extent ($\sim 5\%$), the annihilation cross-section of the stop
NLSP contributes to the total cross-section by the process $\tilde
t_1\,\tilde t_1\to g\,g$.

Next we consider the $\epsilon_1=-0.1$ case (the two middle panels).
The features of the DM allowed regions are similar to the previous case.
The main difference comes from the important enhancement of the Higgs mass
due to the presence of the BMSSM operators. In this case it is possible to
disentangle the $Z$ and the $h$ peaks, since the Higgs-related peak moves to
higher $M_2$ values, due to the increase of the Higgs mass: $m_h=122(150)$
GeV for $\tan\beta=10(3)$. Furthermore, the latter peak is no longer excluded
by chargino searches.

For large values of the $\mu$ parameter, the BMSSM operators
destabilize the scalar potential. The regions above the horizontal
dashed lines in Fig. \ref{dm4} exhibit a metastable electroweak
vacuum, following from the violation of Eq. \eqref{eq:VstabEW} which
requires $\mu\lsim m_A$. Similarly to the situation in the previous
section, points right above the approximate analytical stability line
may still be acceptable, as the stability constraint is somewhat
alleviated by quantum corrections. However, with stop parameters as
specified, computing the tunneling action for the middle and lower
right panels reveals that the limit dictated by Eq. \eqref{eq:VstabEW}
is in fact accurate to better than $15\%$. Hence a significant portion
of the parameter space in the middle and lower right panels is indeed
excluded by stability considerations.

The role of $\epsilon_2$ can be seen from the lower two panels. It
alleviates the stability constraint and slightly increases the Higgs mass.
The $\epsilon_2$-related effect on the Higgs mass is suppressed
by $\tan^2\beta$ and consequently it is much more pronounced in the
$\tan\beta=3$ case than in the $\tan\beta=10$ case.

\section{The EWPT in the BMSSM}
\label{sec:ewpt}
To study the electroweak phase transition, we analyze the finite
temperature effective potential at two-loop order for the light scalar
field. Detailed analyses for the case of the MSSM have been performed
in Refs.~\cite{brignole,Espinosa:1996qw,Carena:1996wj,all}. Here we
focus on the region of parameter space of large $m_A$, for which there
is a single light Higgs field at the phase transition. We review the
main effects found in Ref.~\cite{Blum:2008ym} and study in more detail
the interplay between the relevant parameters. We write the effective
potential at finite temperature as follows:
\beq
V_{\rm eff}(\phi,T)=\frac{T^2\phi^2}{2}
\left[\gamma-B\ln\left(\frac{\phi}{T}\right)\right]
-\frac{m^2}{2}\phi^2-ET\phi^3+\frac{\lambda}{2}\phi^4.
\eeq
The leading contributions to the $\gamma$-term and the $E$-term arise
at one-loop. The $\gamma$-term is further corrected at two-loops by
the so-called $D_{SV}$ and $D_{SS}$ terms \cite{Espinosa:1996qw}.
The $B$-term contains the dominant two-loop corrections in the MSSM,
which arise from the $D_{SSV}$ and $D_{SSS}$ diagrams that contribute
with a logarithmic dependence \cite{Espinosa:1996qw}. For the numerical
analysis we keep, in addition to the SM contributions, all contributions
associated with the light right-handed stop. When considering a light
left-handed stop, the relevant, dominant two-loop (order $g_s^2 h_t^2$
and $h_t^4$) corrections are added as well.

The strength of the phase transition is determined by the order
parameter which, at the critical temperature of the phase transition
$T_c$, is given by
\beq\label{phi/T}
\frac{\phi(T_c)}{T_c} = \frac{E}{2\lambda} + \frac{1}{2}
\left(\frac{E^2}{\lambda^2} + \frac{2 B}{\lambda}\right)^{1/2}
\approx \frac{E}{\lambda} \left(1 + \frac{ \lambda  B}{E^2}\right).
\eeq
We now discuss the various relevant parameters ($E,B,\lambda$) and
their dependence on the MSSM parameters.

The cubic term arises in the regime for which the high temperature
expansion of the bosonic one-loop contribution to the effective
potential is valid. We write
\beq
E = E_{\rm SM} + E_{\rm MSSM}.
\eeq
The SM contribution is given by
\beq
E_{\rm SM}=\frac{2m_W^3+m_Z^3}{6\pi v^3}.
\eeq
The dominant MSSM contribution comes from the stops (we provisionally
neglect the $\tilde t_L-\tilde t_R$ mixing):
\beq
\delta V =-\frac{2 N_c T}{12\pi}
\left(m_{\tilde{t}_{L}}^3(T)+m_{\tilde{t}_{R}}^3(T)\right).
\eeq
The finite temperature stop masses are given by
\beqa
m_{\tilde{t}_{L}}^2(T)&=&m_Q^2+\frac{1}{2}h_t^2 s^2_\beta \phi^2 +
\frac{1}{8} g^2 c_{2\beta} \phi^2 + \alpha_L T^2,\no\\
m_{\tilde{t}_{R}}^2(T)&=&m_U^2+\frac{1}{2}h_t^2
s^2_\beta\phi^2+\alpha_R T^2,
\eeqa
where we neglect contributions from the $U(1)_Y$ gauge bosons
in the field dependent terms. To maximize the value of $E_{\rm MSSM}$,
one would like to take negative values of $m_Q^2$ and $m_U^2$,
\beq\label{mQmU}
m_Q^2 =- \alpha_L T^2, \hspace{1cm} m_U^2 = -\alpha_R T^2,
\eeq
that would cancel the thermal masses, yielding a purely cubic form in
$\phi$ \cite{Carena:1996wj}:
\beq\label{emssmmax}
E_{\rm MSSM}^{\rm max} = \frac{N_c}{3\pi} h_t^3 s^3_\beta.
\eeq
Eq. \eqref{emssmmax} illustrates the effect of the stops and gives, to
leading order, what would be an upper bound on the strength of the
phase transition for the MSSM. It is impossible, however, to make the
selection (\ref{mQmU}) simultaneously for both stops, due to
constraints from the $\rho$ parameter and the experimental bound on
the sbottom mass (through its dependence on $m_Q$).

The two-loop stop contribution to the finite temperature potential,
which can increase $\phi(T_c)$ via its effect on $B$, is given by
\beq
\delta V = -\frac{g_s^2 (N_c^2-1) T^2}{16\pi^2}
\left[ m_{\tilde{t}_{L}}^2(T) \log \frac{2  m_{\tilde{t}_{L}}(T)}{3T}
 +  m_{\tilde{t}_{R}}^2(T) \log \frac{2  m_{\tilde{t}_{R}}(T)}{3T}\right].
\eeq
The effect is maximal when the limit of Eq. (\ref{mQmU}) is realized:
\beq
B^{\rm max}_{\rm stops} = \frac{g_s^2 g_W^2 m_t^2}{\pi^2 m_W^2}.
\eeq
Other logarithmic terms tend to diminish the value of $B$
\cite{Espinosa:1996qw}. Note, however, that the net two-loop
contributions to Eq. (\ref{phi/T}) is the same when considering the
maximal contributions from both stops or from a single one of them.

The strength of the phase transition is further affected by the
quartic coupling $\lambda$. At zero temperature, it is related to the
Higgs mass via $m_h^2 = \lambda_{\rm eff} v^2$, where
\beq\label{mhsq}
m_h^2=m_Z^2 c^2_{2\beta}-16 v^2\epsilon_1\cot\beta
+\frac{3}{4\pi^2} m_t^2 h_t^2 \ln\frac{m_{\tilde{t}_L}
  m_{\tilde{t}_{R}}}{m_t^2}.
\eeq
Eq. (\ref{mhsq}) is valid for large $m_A$ and large $\tan\beta$.  It
includes the leading one-loop corrections. The $\epsilon_2$ dependence
is dropped.  Adding in the leading finite temperature correction, we
have
\beq
\lambda(T)=\lambda_{\rm eff}+\frac{3}{4\pi^2}h_t^4 s^4_\beta \ln2.
\eeq
Thus, we can estimate,
\beq\label{phi/Tc}
\frac{\phi_c}{T_c}=\frac{ v^2( E_{\rm SM} + E_{\rm MSSM})}
{m_h^2 +3 (\ln2) h_t^2 s^2_\beta m_t^2/4\pi^2}
\eeq

We now consider three specific cases. We use different values of the
non-renormalizable contribution ($\epsilon_1$) and the loop
contribution ($m_{\tilde t_L}m_{\tilde t_R}$), but in such a way that
$m_h$ is fixed at the experimental lower bound.

{\bf (a)} $m_h^{\rm tree} = 114$ GeV:\\
To keep $m_h=114$ GeV (and by that minimize the denominator of Eq.
(\ref{phi/Tc})), we need the loop correction of the Higgs boson mass
to vanish, namely \beq\label{zeroloop} m_{\tilde{t}_L}
m_{\tilde{t}_{R}} = m_t^2. \eeq This constrains the possible values of
$ m_{\tilde{t}_{L}}^2(T)$ and $m_{\tilde{t}_{R}}^2(T)$, and gives the
largest possible contribution to $E_{\rm MSSM}$ and to $B$, thus
maximizing the strength of the phase transition.  Smaller values of
$m_{\tilde{t}_L} m_{\tilde{t}_{R}}$ are not allowed due to the
experimental bound on $m_h$. Larger values are allowed but lead to a
decrease in the numerator and an increase in the denominator in Eq.
(\ref{phi/Tc}).

\begin{figure}
\begin{center}
\includegraphics[width=8.5cm]{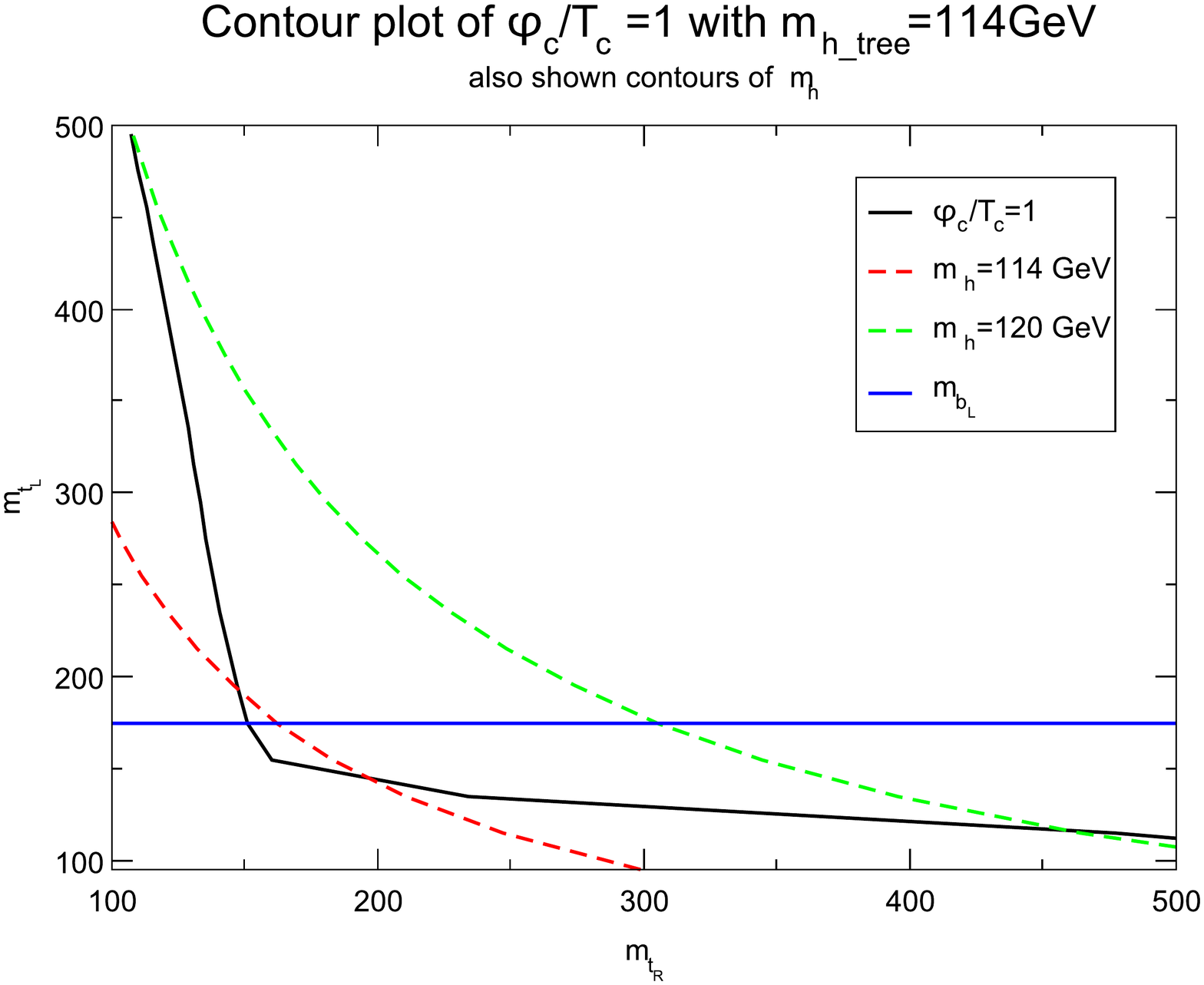}\hspace{-0.8cm}
\includegraphics[width=8.5cm]{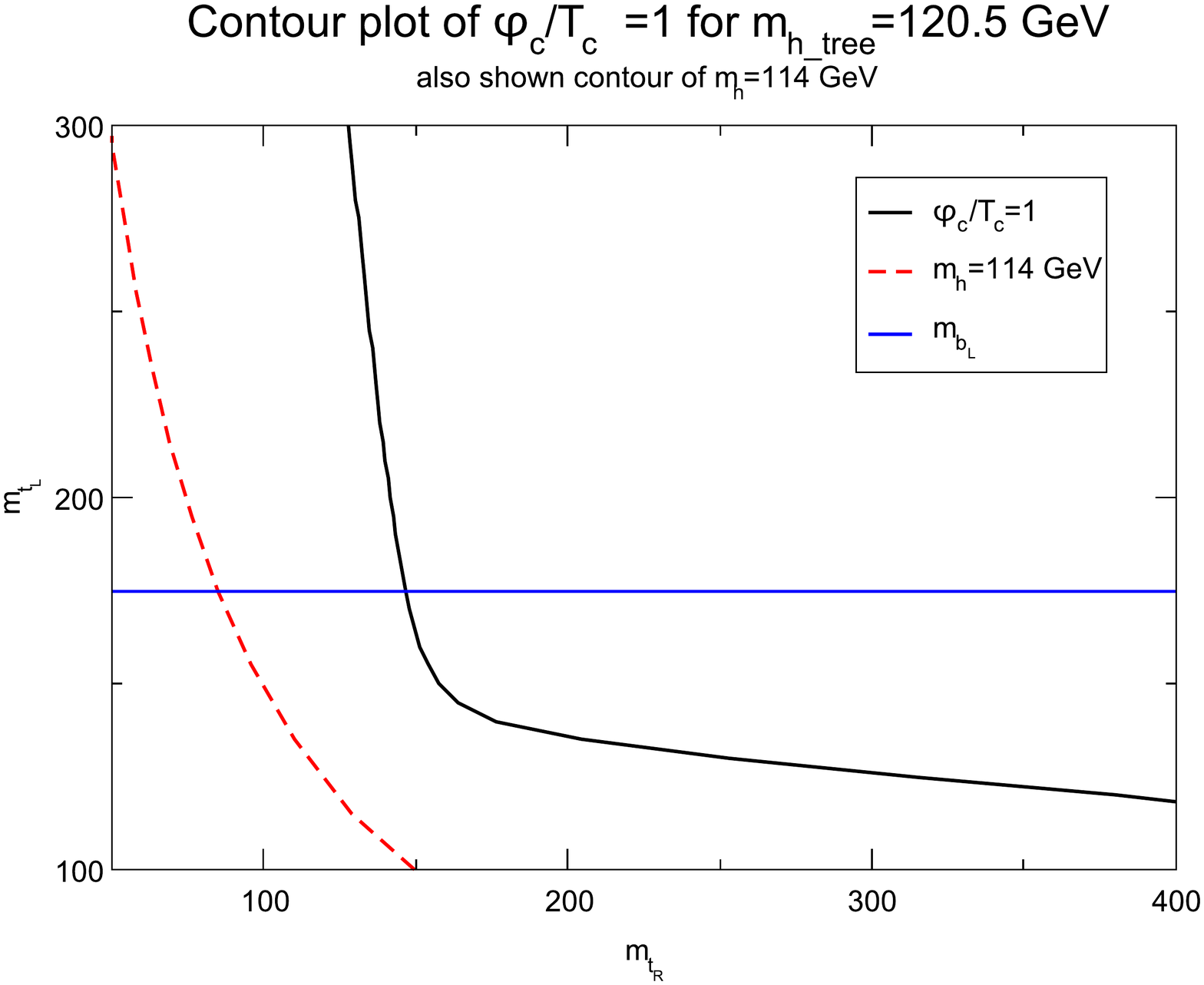}
\end{center}
\vspace{-1.1cm} \caption{Contour of $\phi_c/T_c =1$ (solid black
  curve), and contours of $m_h$ (dashed curves) for $\tan\beta=5$, and
  $\epsilon_1=-0.065$ (left panel) or $-0.08$ (right panel). Also
  shown is the constraint arising from the lower bound on $m_{\tilde
  b_L}$ (solid blue line).}
\label{casea}
\end{figure}

The left panel of Fig. \ref{casea} shows the contour of
${\phi_c}/{T_c}=1$ and the contour of $m_h=114$ in the
$m_{\tilde{t}_{L}}$-$m_{\tilde{t}_{R}}$ plane. We also show the curve
of $m_{\tilde{t}_{L}} \gsim 175$ GeV, corresponding to the
experimental lower bound on $m_{\tilde b_L}$. We use $\tan\beta=5$ and
$\epsilon_1=-0.065$, so that we obtain $m_h^{\rm tree}=114$ GeV. The
$m_h$ curve then corresponds to Eq.  (\ref{zeroloop}). The plot shows
the approximate symmetry in this case between $m_{\tilde{t}_{L}}$ and
$m_{\tilde{t}_{R}}$, which is broken only by effects arising from
D-terms. The allowed region is to the left of the ${\phi_c}/{T_c}=1$
curve and above the $m_h=114$ GeV curve.

{\bf (b)}  $m_h^{\rm tree} > 114$ GeV:\\
Here we take values of $\epsilon_1$ such that $m_h^{\rm tree}> 114$
GeV. In this case, $m_{\tilde{t}_{L}} m_{\tilde{t}_{R}}<m_t^2$ is
allowed. Such values further increase $E_{\rm MSSM}$ with respect to
the values presented in case {\bf (a)}, and the allowed region for a
strong enough phase transition is increased, although here the sbottom
mass constraint eliminates a large portion of this region. The right
panel of Fig. \ref{casea} shows the ${\phi_c}/{T_c}=1$ the $m_h = 114$
GeV contours for this case.

{\bf (c)}  $m_h^{\rm tree} < 114$ GeV:\\
This is reminiscent of the usual MSSM results for the electroweak
phase transition, in which one requires $m_{\tilde{t}_{L}}$ to be
large enough to satisfy the $m_h^{\rm exp}$ value. The effect on
$E_{\rm MSSM}$ is to effectively screen the contribution from the
left-handed stop, thus reducing the cubic term to be smaller than half
its maximum value and, furthermore, reducing the value of $B$.
Variations in the value of $m_Q$ produce only small variations in the
value of $m_U$, as the main effect of increasing $m_Q$ is to increase
the one-loop value of the Higgs boson mass, thus reducing the strength
of the phase transition. To compensate for that, a slightly smaller
value of $m_{\tilde{t}_{R}}$ must be taken to increase $E_{\rm MSSM}$.

In the region of the parameter space where the EWPT is strong enough,
a minimum where colour is broken might develop \cite{Carena:1996wj}.
If the temperature where this minimum becomes as low as the potential
at the origin, $T_c^U$, is higher than the critical temperature for
the EWPT, $T_c^\phi$, then the Universe is likely to end up in the
colour breaking minimum. Thus, this region is excluded
\cite{Cline:1999wi}. When the two stops are light enough, $T_c^\phi$
is safely higher than $T_c^U$. However, when we consider higher and
higher values of $m_{\tilde{t}_{L}}$, and correspondingly, to keep the
EWPT strong enough, lower and lower values of $m_{\tilde{t}_{R}}$, the
closer do $T_c^\phi$ and $T_c^U$ get to each other, until, at some
critical values of $(m_{\tilde{t}_{L}}^c,m_{\tilde{t}_{R}}^c)$, we
reach $T_c^{\phi}=T_{c}^{U}$.

Up to this point, we fixed the values of $\epsilon_1$ and $\tan\beta$
and obtained the allowed regions in the $[m_{\tilde t_R},m_{\tilde
  t_L}]$ plane. We learned that the presence of $\epsilon_1={\cal
  O}(-0.1)$ opens new regions in this plane where both the $m_h$
constraint and the EWPT constraint are satisfied. We now turn our
attention to the dependence on the other relevant parameters. More
concretely, we fix $m_{\tilde t_L} =200$ GeV, which is close to the
lowest value allowed by the $\delta\rho$ parameter, and obtain the
dependence of $\phi_c/T_c$ and of $m_h$ on $\tan\beta,
m_{\tilde{t}_R}$, and $\epsilon_1$.

\begin{figure}[ht!]
\begin{center}
\includegraphics[width=8.5cm]{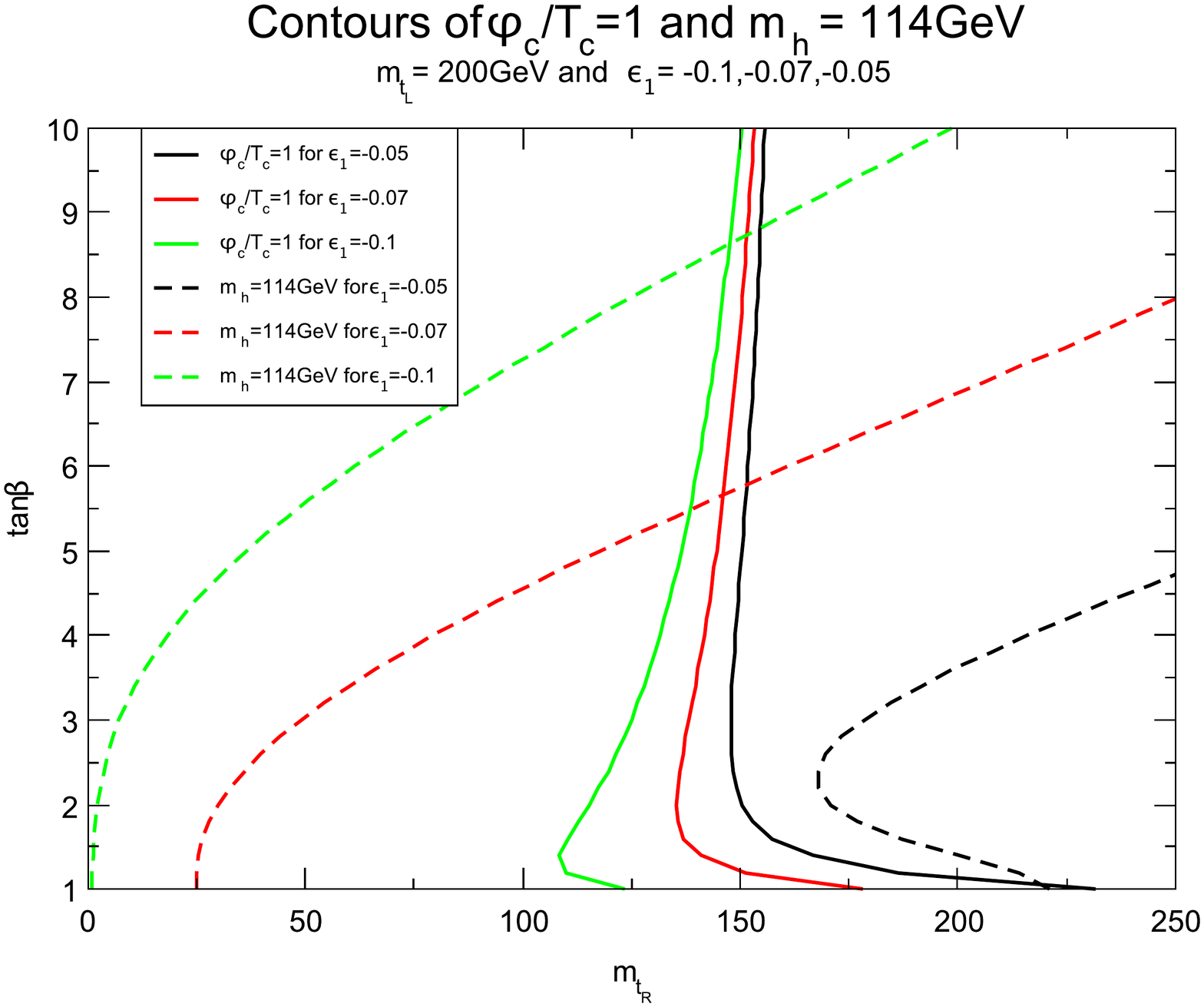}\hspace{-0.8cm}
\includegraphics[width=8.5cm]{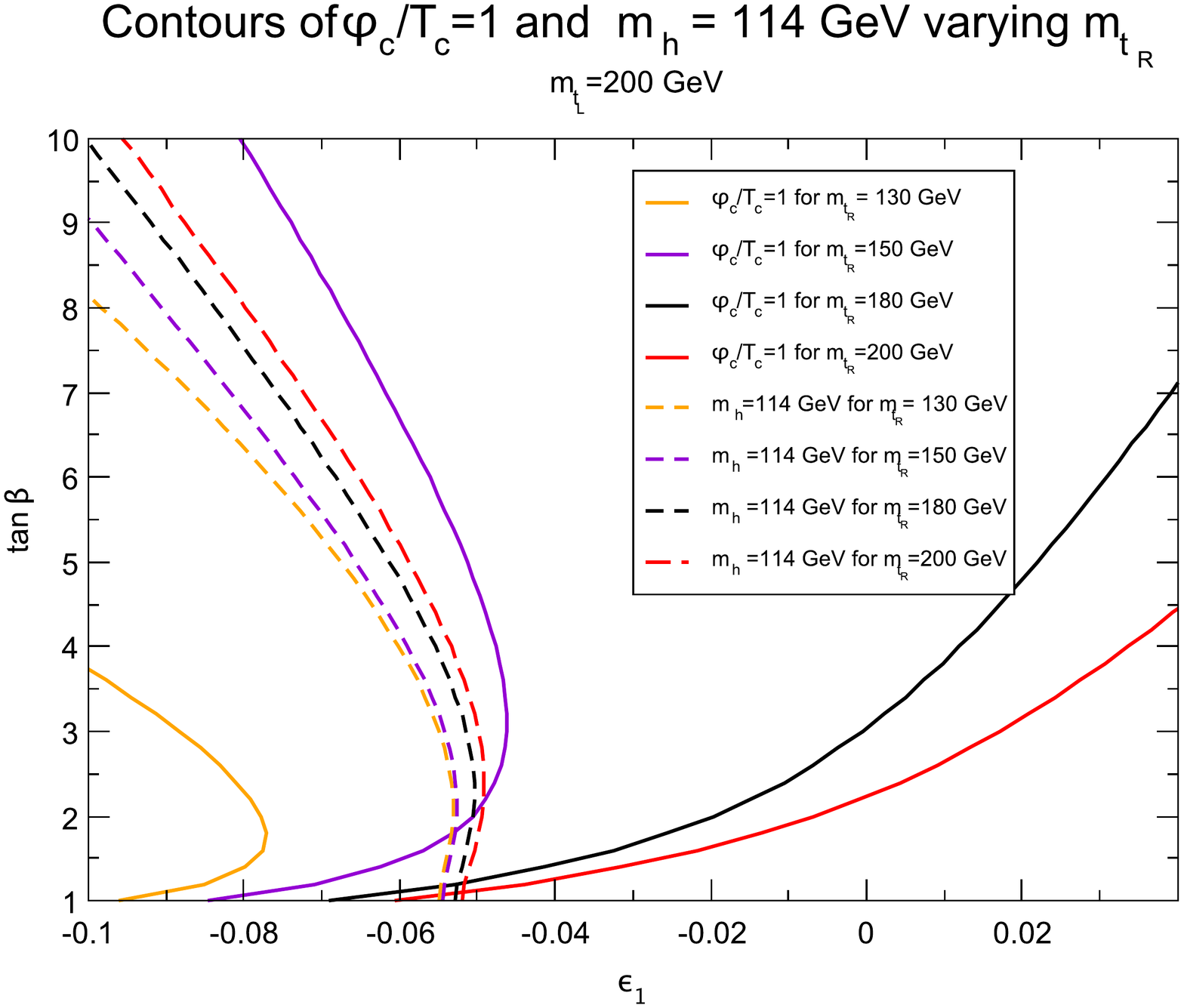}\hspace{-1cm}
\includegraphics[width=8.5cm]{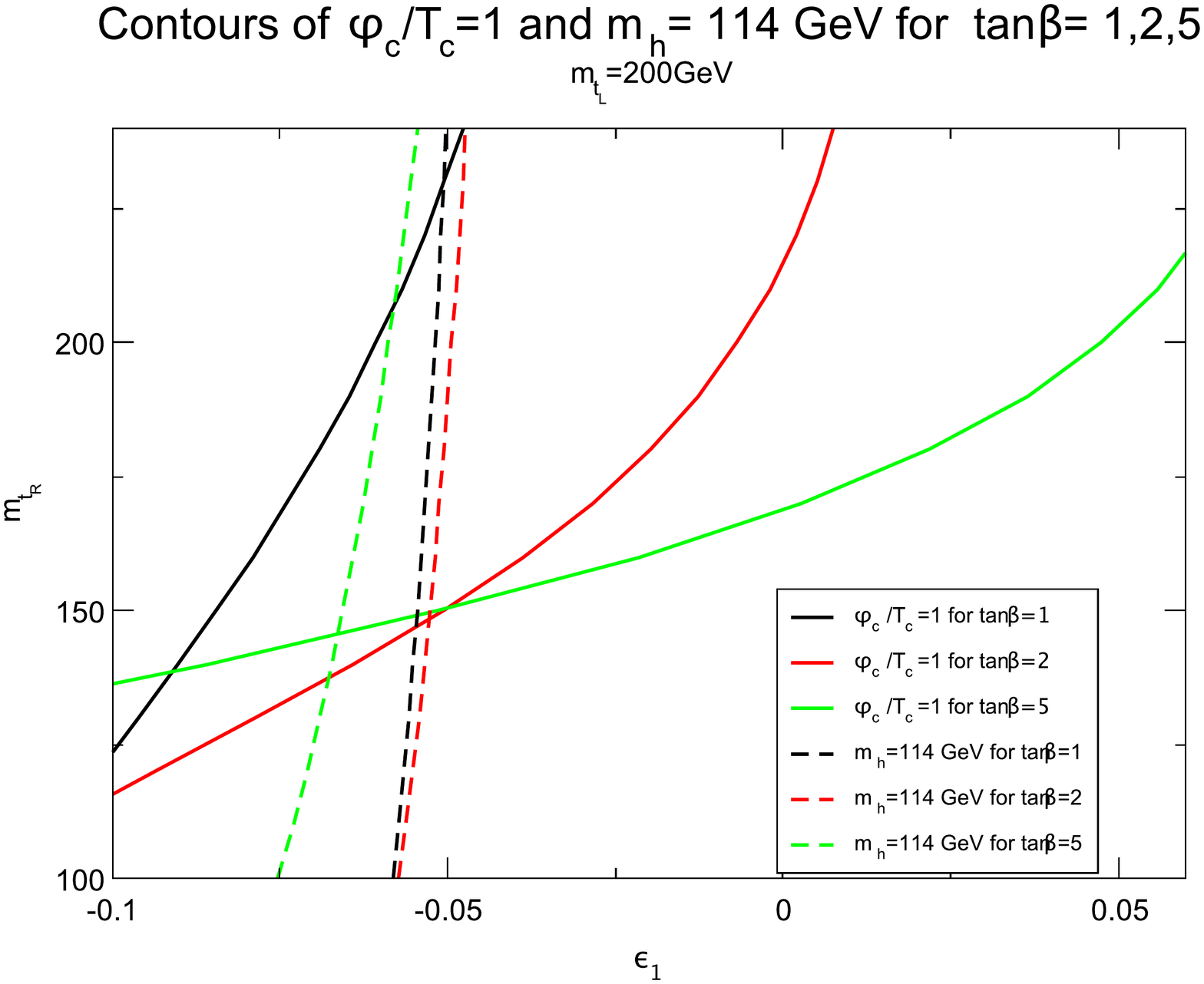}
\end{center}
\vspace{-1.1cm}
\caption{The dependencies of the Higgs mass and the EWPT on
  $\tan\beta$, $m_{\tilde{t}_{R}}$, and $\epsilon_1$. We fix
  $m_{\tilde{t}_{L}}=200$ GeV. Solid lines indicate contours of
  $\phi_c/T_c =1$ and dashed lines contours of $m_h=114$ GeV. Each
  same-colour pair of (solid and dashed) lines corresponds to the
  same set of parameters.}
\label{BAU2}
\end{figure}

Our results are presented in Figs. \ref{BAU2} as contours of
$\phi_c/T_c$ and of $m_h$ in various parameter planes. In the top left
panel we make discrete choices of $\epsilon_1$, and present the
results in the [$m_{\tilde t_R},\tan\beta$] plane. For each value of
$\epsilon_1$, the allowed region is to the right of the $m_h=114$ GeV
and to the left of the $\phi_c/T_c=1$ curve. In the top right panel we
make discrete choices of $m_{\tilde t_R}$ and present the results in
the [$\epsilon_1,\tan\beta$] plane. For each value of $m_{\tilde
  t_R}$, the allowed region is to the left of the $m_h=114$ GeV and to
the right of the $\phi_c/T_c=1$ curve. In the lower panel we make
discrete choices of $\tan\beta$ and present the results in the
[$\epsilon_1,m_{\tilde t_R}$] plane. For each value of $\tan\beta$,
the allowed region is to the left of the $m_h=114$ GeV and below and
to the right of the $\phi_c/T_c=1$ curve. We learn that small values
of $\tan\beta$ are allowed and, furthermore, values of
$m_{\tilde{t}_R}> m_t$ can simultaneously satisfy the requirement for
baryogenesis and the Higgs boson mass bound.

To summarize the results depicted in Figs.~\ref{casea} and~\ref{BAU2},
the BMSSM framework introduces the following significant differences
with respect to the standard MSSM case:
\begin{enumerate}
\item The Higgs boson mass constraint decouples from the value of
  $m_Q$ in some regions of parameter space. Consequently, a very light
  left-handed stop can contribute significantly to the cubic term in
  the effective potential. Explicitly, values of $m_{\tilde{t}_{L}}
  \lsim 175$ GeV ($m_Q \sim 0$) are now allowed, compared to the
  values of $m_Q \sim 2-3$ TeV, that are required in the MSSM.
\item The lowest allowed value for $m_{\tilde{t}_L}$ arises from
  electroweak precision measurements and from direct searches. For the
  MSSM, these constraints are superseded by requiring $m_h>114$GeV.
\item In the MSSM, a large value of left-right mixing in the stop
  sector is strongly favored. This is not necessary within the BMSSM.
  The actual consequences of left-right mixing in the stop sector are
  quite similar between the BMSSM and the MSSM: The Higgs boson mass
  is increased, which, in turn, weakens the phase transition (although
  this can be compensated by changing $\epsilon_1$). Moreover, the
  contributions to the cubic term in the potential from the stops are
  screened, again weakening the phase transition.
\item Larger values of $m_{\tilde{t}_{R}}$ are allowed. Actually, the
  experimental bounds on the masses of the left handed squarks
  determine the largest possible mass of the right-handed stop.
\item Smaller values of $\tan\beta$ are now feasible.
\end{enumerate}

We learn that the greater freedom in the values of the different
parameters that affect the Higgs boson mass -- $\tan\beta$,
$m_{\tilde{t}_{L}}$, $m_{\tilde{t}_{R}}$ -- implies that the allowed
regions are similar to those identified in the early analyses of the
EWPT \cite{brignole}.

Two final comments are in order. First, note that the inclusion of
non-renormalizable operators of dimension six could modify the
results. In particular, it is clear that for large $m_A$, such terms
can induce additional contributions like those identified in Ref.
\cite{Grojean:2004xa}. Vacuum stability considerations must be
carefully taken into account in this case.

Second, as pointed out in Refs. \cite{Dine:2007xi,Blum:2008ym}, the
non-renormalizable operators can induce new sources of CP violation in
the scalar Higgs sector, that would modify the production mechanism of
the baryon asymmetry of the Universe. These new sources could provide
a relief to the tension between the large phases required to produce
the BAU and the contributions to the electric dipole moments.
Additionally, even in the absence of CP violation in the scalar Higgs
sector, modifications to the EDMs arise from the new interaction terms
between the Higgs bosons and the higgsinos \cite{PRS}. These issues will be
further studied elsewhere.

\section{Conclusions}
\label{sec:con}
The main motivation to add non-renormalizable operators to the MSSM
Higgs sector is to reduce the fine-tuning that is required by the
lower bound on the Higgs mass. We have shown that, in addition, these
operators have implications for cosmology that are very welcome:
Regions of the supersymmetric parameter space that are favored by the
dark matter constraints and by the requirement for a strong
first-order electroweak phase transition, but are excluded within the
MSSM, become viable within the BMSSM.

As concerns dark matter, a particularly important feature of the BMSSM
is that the bulk region, for which the LSP is mostly bino-like and the
sfermions are relatively light, can provide the adequate contribution
to the energy density of the Universe while still satisfying the
collider constraints on the Higgs boson as well as on the
supersymmetric particle spectrum. Light stops co-annihilating with the
LSP could have been active players in driving the dark matter relic
density to its present value. It is also possible that nearly resonant
LSP annihilation proceeded through exchange of the lightest Higgs
particle itself.

If light stops are indeed found in upcoming experiments, large
BMSSM corrections will be implied. In this scenario, parameter regions
where $\mu$ is large (exhibiting some heavy neutralinos and charginos)
will be significantly constrained by the requirement of vacuum
stability.

As concerns the electroweak phase transition, the BMSSM has a dramatic
effect when determining the range of parameters for which the phase
transition is sufficiently strong to suppress sphaleron transitions in
the broken phase. The fact that large stop-related radiative
corrections to the Higgs mass are not required, allows light stop
degrees of freedom to affect the dynamics of the phase transition by
enhancing their contributions to the magnitude of the order parameter
at the critical temperature.

\subsection*{Note added:} Upon completion of this work,
a related paper \cite{Berg:2009mq} has appeared.
Where the two analyses overlap, we confirm their results.

\section*{Acknowledgements}
We thank Cedric Delaunay for useful discussions. Results from this
work were presented at the Planck 2009 conference in May 2009. The
work of YN is supported by the Israel Science Foundation (ISF) under
grant No.~377/07, the United States-Israel Binational Science
Foundation (BSF), Jerusalem, the German-Israeli foundation for
scientific research and development (GIF), and the Minerva Foundation.
The work of ML was supported in part by Colciencias under contract
11150333018739.  NB thanks an ESR position of the EU project RTN
MRTN-CT-2006-035505 HEPTools. The work of NB and ML was supported in
part by the Ecos-Nord program.


\end{document}